\documentclass[instruments,article,authorversion,moreauthors,LaTeX,dvi2pdf,10pt,a4paper,tikz]{mdpi}
\usepackage{multirow}
\usepackage{graphicx}
\usepackage{graphbox}
\usepackage{lscape}
\usepackage{verbatim}
\usepackage{rotating}
\usepackage{tabularx, booktabs}

\firstpage{1} 

\Title{Secondary beams at high-intensity electron accelerator facilities}

\Author{Marco Battaglieri$^{1}$, Andrea Bianconi$^{2,3}$, Mariangela Bond\'i$^{4}$, Raffaella De Vita$^{1}$, Antonino Fulci$^{4,5}$*, Giulia Gosta$^{2}$, Stefano Grazzi$^{1,5}$, Hyon-Suk Jo$^{6}$,
Changhui Lee$^{6}$, Giuseppe Mandaglio$^{4,5}$, Valerio Mascagna$^{2,3}$, Tetiana Nagorna$^{1}$, Alessandro Pilloni$^{4,5}$,
Marco Spreafico$^{1,7}$, Luca J Tagliapietra$^{8}$, Luca Venturelli$^{2,3}$ and Tommaso Vittorini$^{1,7}$}

\AuthorNames{Marco Battaglieri, Mariangela Bond\'i, Raffaella De Vita, Antonino Fulci, Giulia Gosta, Stefano Grazzi, Hyon-Suk Jo, Changhui Lee, Tetiana Nagorna, Alessandro Pilloni, Marco Spreafico, Luca J Tagliapietra, Tommaso Vittorini}

\address{%
$^{1}$ \quad Istituto Nazionale di Fisica Nucleare, Sezione di Genova, 16146 Genova, Italy\\
$^{2}$ \quad Istituto Nazionale di Fisica Nucleare, Sezione di Pavia, 27100 Pavia, Italy\\
$^{3}$ \quad Università degli Studi di Brescia, 25123 Brescia, Italy\\
$^{4}$ \quad Istituto Nazionale di Fisica Nucleare, Sezione di Catania, 95125 Catania, Italy\\
$^{5}$ \quad Universit\`a degli Studi di Messina, 98166 Messina, Italy\\
$^{6}$ \quad Kyungpook National University, Daegu 41566, Republic of Korea\\
$^{7}$ \quad Universit\'a degli Studi di Genova, 16126 Genova, Italy\\
$^{8}$ \quad NEVNUCLAB, 123 W Nye Lane, Carson City, Nevada, USA.\\
}

\corres{Corresponding author: antonino.fulci@unime.it}

\abstract{The interaction of a high-current $O$(100~\textmu~A), medium energy $O$(10~GeV) electron beam with a thick target $O$(1m) produces an overwhelming shower of standard model particles in addition to hypothetical Light Dark Matter particles.While most of the radiation (gamma, electron/positron) is contained in the thick target, deep penetrating particles (muons, neutrinos, and light dark matter particles) propagate over a long distance, producing high-intensity secondary beams.
Using sophisticated Monte Carlo simulations based on FLUKA and GEANT4, we explored the characteristics of secondary muons and neutrinos and (hypothetical) dark scalar particles produced by the interaction of Jefferson Lab 11~GeV intense electron beam with the experimental Hall-A beam dump. Considering the possible beam energy upgrade, this study was repeated for a 22~GeV CEBAF beam.}


\begin{document}

\setcounter{section}{0}

\section{Introduction}
High-intensity particle beams represent one of the current frontiers of discovery in particle 
and nuclear physics. High-intensity proton beams are routinely used to generate 
secondary beams of particles such as neutrinos and muons that could be used to extend 
the exploration of matter with new and different probes.

High-current ($\sim$100~\textmu A), medium-energy (1~GeV--10~GeV), 
continuous-wave electron beams with a delivered large integrated charge 
($\sim$1000~C/y) can be also used to generate secondary beams.  
In fixed-target experiments, after the interaction with a  thin target, the beam is dumped 
on a block of material where electrons produce showers, degrading the initial energy 
down to values at which ionization and excitation of atoms dominates. If the primary 
beam's initial energy is higher than the pion production threshold, hadronic interaction 
and electromagnetic processes contribute to the production of a sizable number of 
secondary particles that may re-interact or escape from the dump. The beam dump (BD) is 
usually surrounded by heavy shielding (e.g., a thick concrete vault) to minimize the 
escaping radiation. Nevertheless, a significant flux of neutrons, muons, and neutrinos 
propagate through the shielding, making intense secondary beams that may provide an 
opportunistic extension of investigations performed with the primary electromagnetic 
probe. According to recent theoretical studies, the interaction of an intense electron beam 
with the beam dump could also be a source of a light dark matter (LDM) particle 
beam~\cite{Battaglieri:2017aum}. LDM particles are viable candidates to explain 
gravitational anomalies, extending the current set of elementary particles and interactions 
beyond the standard model (BSM).     

The electron's prevalent electromagnetic interaction represents an alternative and complementary method of producing intense secondary beams that differ from hadronic-initiated reactions for the energy spectrum, the spatial dispersion, and the associated background. 
Using FLUKA and GEANT4, the state-of-the-art simulation tools widely used in 
high-energy and nuclear physics, we studied and characterized the secondary muons, 
neutrinos, and (hypothetical) LDM beams produced at the Jefferson Lab via the interaction 
of the primary electron beam with the experimental Hall-A BD. A similar study with 
higher energy lepton beams has been performed for the future ILC facility 
\cite{SAKAKI2023168144}.

The paper is organized as follows: in Section~\ref{sec:JLAB}, some details about the  
Thomas Jefferson National Accelerator Facility (Jefferson Lab or JLab) are reported. In 
Section~\ref{Sect_simul_framework}, the simulation framework used to derive the 
secondary beams is described. Sections \ref{sec:SecondaryMu}--\ref{sec:DarkScalar} 
report the expected characteristics of muon, neutrino, and LDM secondary beams. The 
conclusions and future outlook are reported in the last section.

\section{The Thomas Jefferson National Accelerator Facility}
\label{sec:JLAB}
The Jefferson Lab is a US Department of Energy laboratory located in Newport News, 
Virginia. JLab hosts the Continuous Electron Beam Accelerator Facility (CEBAF), a 
continuous wave (CW) electron accelerator, made by two 1-GeV LINACs and recirculating 
arcs to achieve, in five passes, the maximum energy. The machine started operations in 
1994, delivering a 4~GeV beam, which was soon upgraded to 6~GeV and later to 12~GeV. 
At present, four experimental halls can receive, simultaneously, a primary 11~GeV 
electron beam (Hall-A, -B, and -C) and up to 12~GeV electron beam (Hall-D, then 
converted to a secondary photon beam)  to conduct scattering experiments on nucleons 
and nuclei. The physics program includes the study of the hadron spectrum,  nucleon 
structure, nuclear interaction, and BSM searches. The excellent quality of the polarized 
electron beam allows one to run high-precision parity-violation experiments that use the 
interference between electromagnetic and weak interaction to study the properties of 
quarks inside hadrons and nuclear matter.\
Hall-A and Hall-C are equipped with high-precision magnetic spectrometers. The 
detector's small acceptance requires high current (1--150~\textmu A) on target to reach the 
typical luminosity of 10$^{39}$~cm$^{-2} \text{s}^{-1}$. The high-current operations make 
Hall-A BD the ideal source of secondary beams at Jefferson Lab. The current BD 
configuration limits the maximum power to $<$1~MW corresponding to 90~\textmu A 
current at 11~GeV beam energy or higher current at lower energy.
High-current experiments with long durations are planned for Hall-A in the next decade, 
while the number of running days per year for Hall-C will be more sparse.
Hall-B and Hall-D host two large-acceptance spectrometers (CLAS12 and GlueX) based on 
a toroidal (CLAS12) and solenoidal (GlueX) magnetic field. The almost 4$\pi$ acceptance 
limits the current to hundreds of nA in Hall-B, and to a few \textmu A in  Hall-D's 
radiator (to generate a Bremsstrahlung real-photon beam). Dumps installed in these two 
halls are limited to a power of $\sim$100~kW, reducing the intensity of the incoming 
primary beam to values unsuitable for generating intense secondary beams. For this 
reason, we focused this study on the Hall-A beam dump only.

Currently, a study to increase the maximum beam energy of the CEBAF accelerator 
complex is underway. Taking advantage of progress in accelerator technologies, it will be 
possible to extend the energy reach of the CEBAF accelerator up to 
22~GeV~\cite{accardi2023strong} within the current tunnel footprint and reusing the 
existing superconducting radio frequency (SRF) cavity system. Using the fixed-field 
alternating-gradient (FFA) technique, it will be possible to increase the number of passes 
through the accelerating cavities by reusing the same recirculating arcs. Considering the 
possibility of this upgrade, we studied the characteristics of secondary beams in the two 
configurations: the existing 11~GeV primary electron beam energy and a future 22~GeV.

\section{The Simulation Framework}
\label{Sect_simul_framework}
The interaction of the 11\,GeV (22\,GeV) primary electron beam with the Hall-A BD and 
subsequent transportation of the secondary muon, neutrino, and LDM  beams was studied 
by Monte Carlo simulations using FLUKA~\cite{Bohlen:2014buj,Ferrari:2005zk} and 
GEANT4~\cite{Agostinelli:2002hh} toolkits. The same simulation framework was 
successfully used to describe the results of the BDX-HODO~\cite{BATTAGLIERI2019116} 
and BDX-MINI \cite{Battaglieri:2022dcy} experiments at Jefferson Lab.  


\subsection{FLUKA}
 \label{sect_FLUKA}
FLUKA~\cite{fluka-cern, fluka-cern-art, fluka-cern-art2, fluka-neutrino} version 4-3.1 was 
used to simulate the interaction of the primary electron beam with the Hall-A beam dump 
and the propagation of muons and neutrinos through concrete and dirt to reach  a 
hypothetical downstream detector.  
The Hall-A beam dump geometry and materials were implemented according to the prescriptions of JLab Radiation Control Group~\cite{Kharas}. 

\begin{figure}[H]
 \centering
    \includegraphics[width=.99\textwidth]{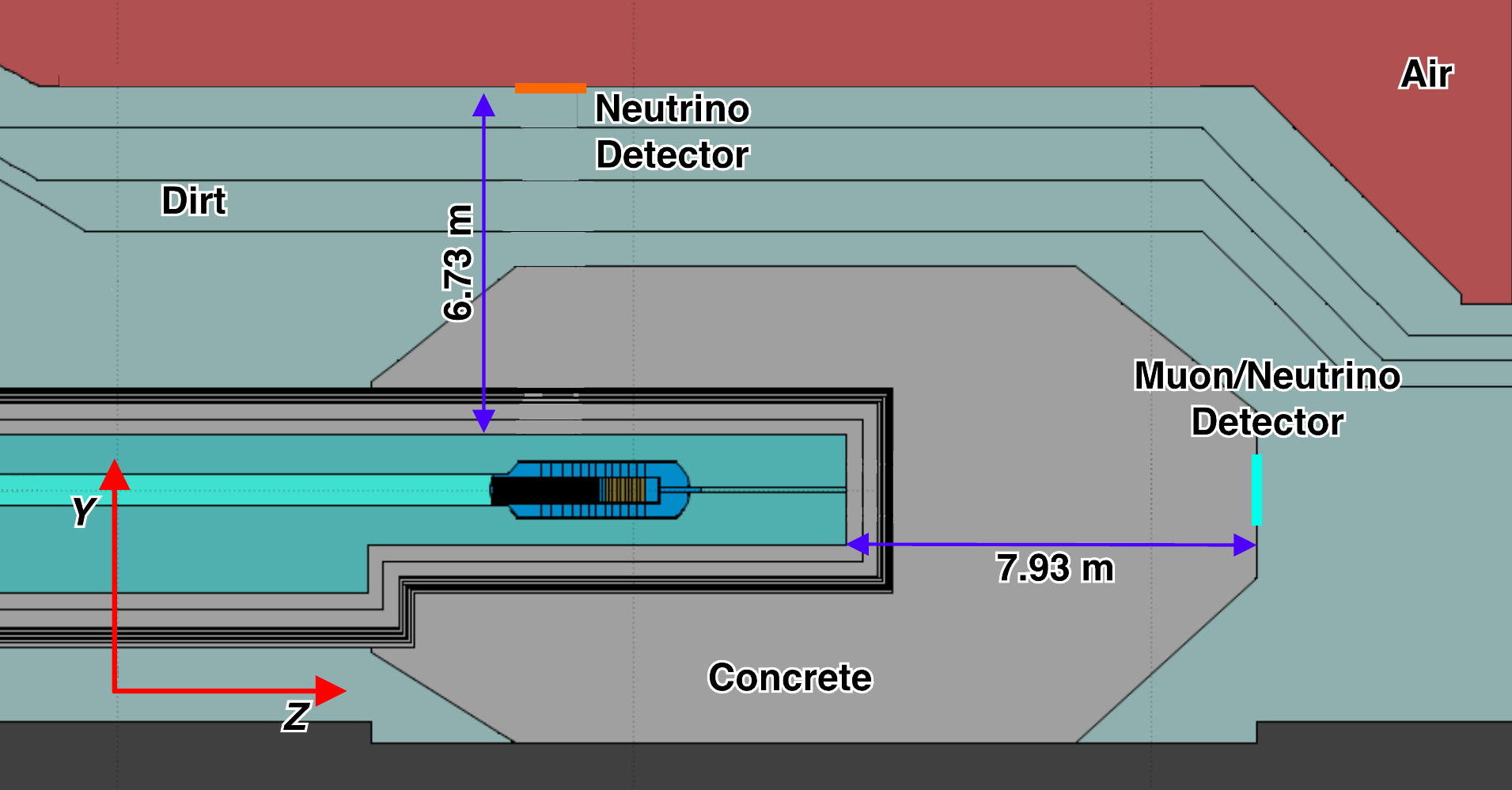}
    \caption{Scale side cross section of the Hall-A beam dump geometry with the two 
    flux-detectors used in the simulations to evaluate the flux of secondary particles. 
    Perpendicular to the beam dump, in orange, is shown the flux detector corresponding to 
    the location of an hypothetical neutrino detector. Aluminum disks of beam dump inner 
    core are shown in yellow. Downstream of the beam dump, immediately after the 
    concrete vault, a hypothetical muon/neutrino detector's location is shown in light blue.}
    \label{fig:geometry_YZ}
\end{figure}

The beam dump consists of approximately 80 aluminum disks, each with a diameter of 
roughly 40~cm.  The disk thickness progressively increases from 1~cm to 2~cm, spanning a 
cumulative length of about 200~cm. Downstream of the disks, there is an aluminum 
cylinder, measuring 50~cm in diameter and approximately 100~cm in length. To ensure 
optimal temperature control,  disks and cylinders are thermalized using a water-cooling 
circuit. To enhance the radiation shielding capabilities, the beam dump is surrounded by 
$\sim$8~m of concrete in the longitudinal direction (where the high-energy secondary 
particles are produced) and about $\sim$3~m of concrete in the transverse direction. 
Furthermore, the entire setup is covered by $\sim$4 m of overburden.
The BD, the beam transport line, and the surrounding concrete vault are shown (to scale) 
in Figure~\ref{fig:geometry_YZ}.

\begin{figure}[H]
     \centering
    \includegraphics[scale=0.2]{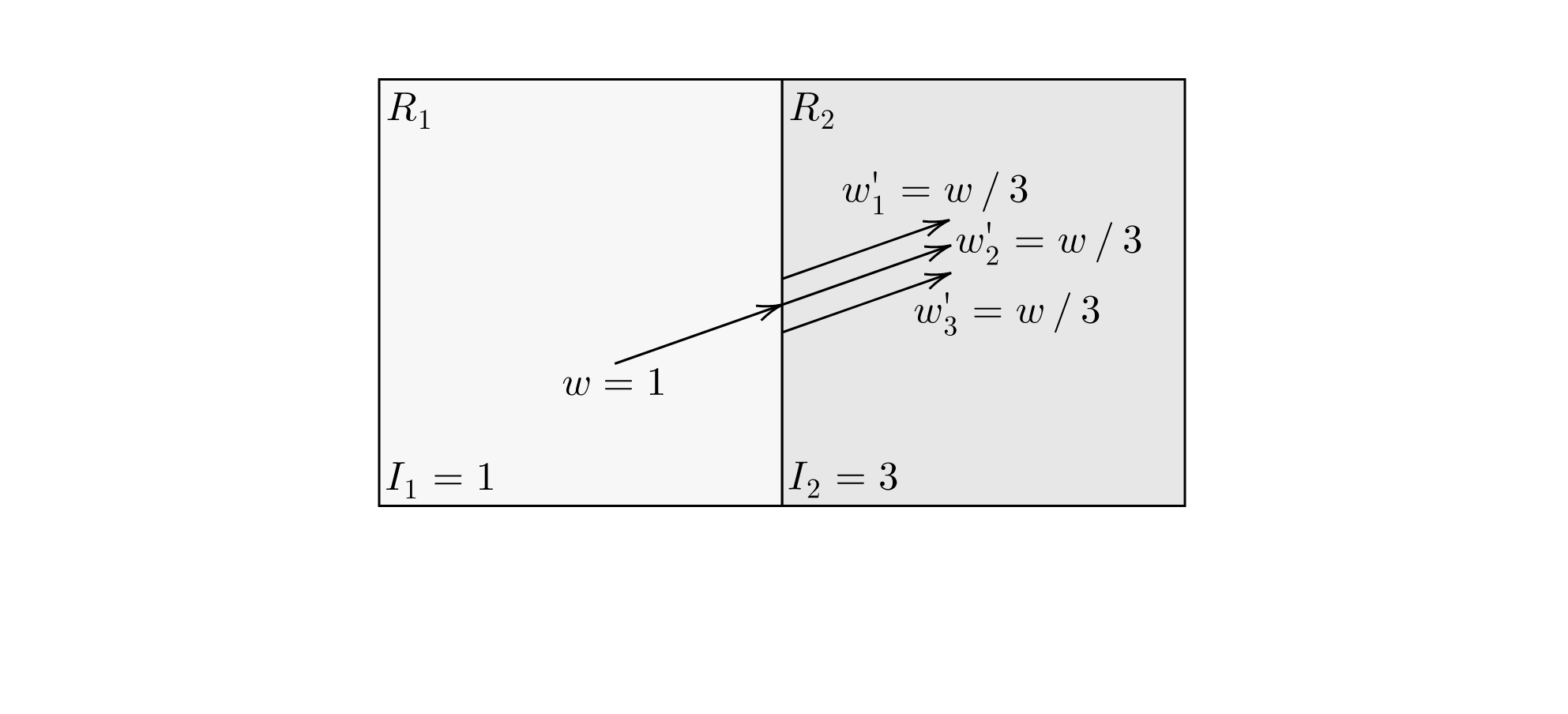}
    \caption{Surface splitting biasing mechanism scheme. A particle passing into a region with higher importance is split and given a weight equal to its initial weight divided by the importance of the current region.}
    \label{fig:surface-splitting}
\end{figure}

The input parameters used to run the program include all physics processes and a tuned 
set of biasing weights. As a reference, we considered  a run time of 1 year corresponding to 
$\sim$$10^{22}$ EOT. Simulations corresponding to such high statistics would require an 
unpractical time with the available computing resources. To speed up the running time 
while preserving accuracy, biasing techniques were used  to effectively reach the target 
statistics.
Two biasing techniques provided by FLUKA were employed (an overview of biasing techniques in Monte Carlo simulations is described in Ref. \cite{Ferrari:2005zk}). The first, known as \textit{surface splitting 
}, involves splitting a particle when it crosses two regions of increasing importance, as 
depicted in Figure~\ref{fig:surface-splitting}. The second technique, referred to as 
\textit{interaction length biasing}, involves a reduction in photon--nucleus interaction 
length, thereby increasing the number of particles produced, especially muons.

The user FLUKA routines, written in FORTRAN and C++ programming languages, were 
used to generate a custom output file as a ROOT TTree~\cite{BRUN199781}. This 
approach offers several advantages. Firstly, it facilitates the parallelization of the 
simulation across multiple CPUs, as the data can be easily merged at the end of the 
simulations. Secondly, it enhances flexibility in post-simulation analysis of results.
Unlike the standard FLUKA scorers, which require predefined input settings, the 
\texttt{TTree} format allows data post-processing with no need to run the same 
simulations multiple times.
Whenever a particle crosses the boundary between two selected regions, the following 
information is saved: crossing surface identifier, particle ID, statistical weight, total energy 
and momentum, crossing position vertex, direction (represented as direction cosine), 
parent particle ID, parent particle energy, production vertex, and production process 
identifier. The stored information is subsequently processed using dedicated ROOT 
scripts, written in Python, leveraging the pyROOT interface~\cite{pyROOT}.

\subsection{GEANT4}
\label{sec:GEMC}

\begin{figure}[H]
   \centering
    \includegraphics[width=.99\textwidth]{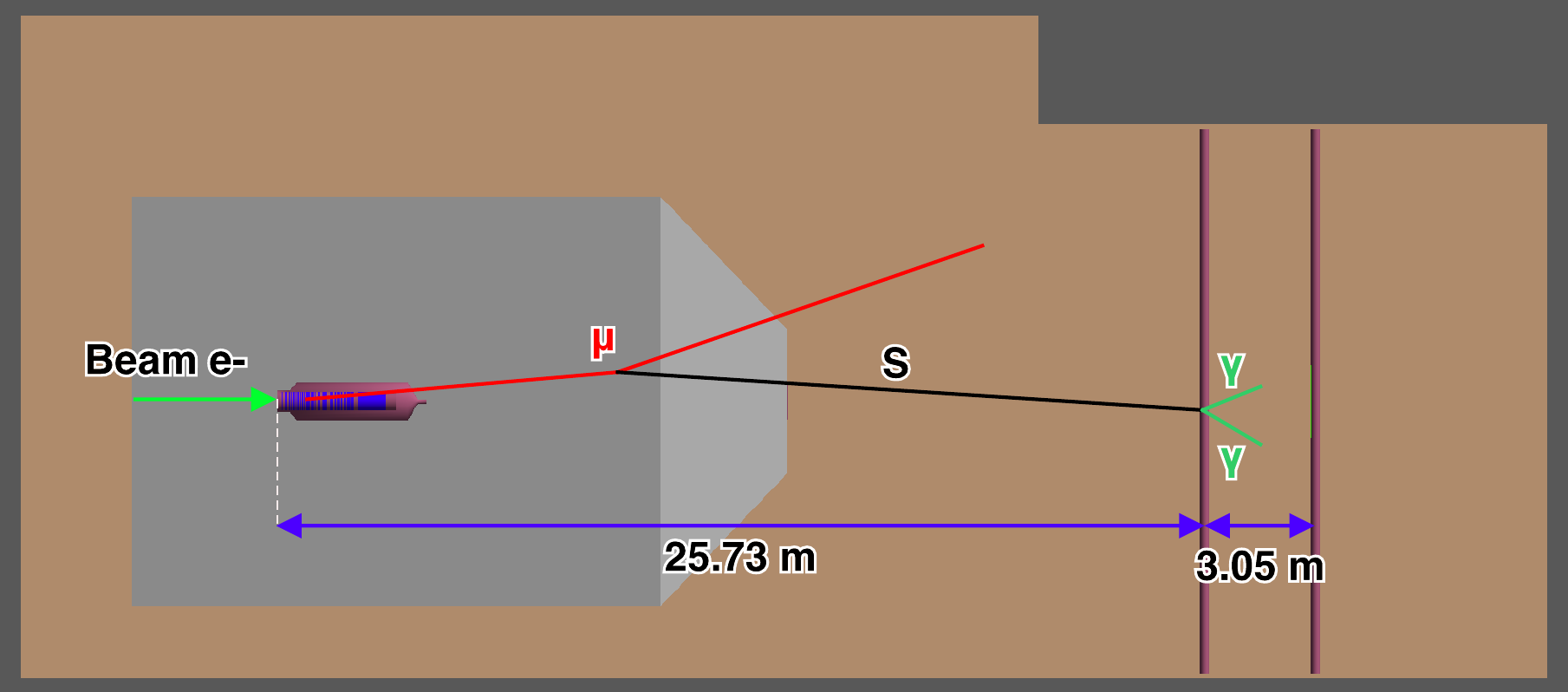}
    \caption{Hall-A 
 BD  and surrounding dirt implemented in GEMC. The BD vessel, shown in purple, contains Al foils, in blue.
    The concrete vault is shown in gray, while the dirt is shown in brown. Two existing 
    $10"$ pipes installed $\sim$26~m and $\sim$29~m downstream of the dump 
    \cite{battaglieri2019dark} are shown in purple. A scheme of dark scalar production and 
    decay is also presented. }
    \label{fig:GEMC_geometry}
\end{figure}

LDM can be produced by the interaction of standard model particles with ordinary matter. 
In this paper, we only consider DM particles produced by the interaction of the secondary 
muon beam with the BD and surrounding materials. (LDM produced in the 
 direct interaction of the primary electron beam with the BD was studied and reported in 
 Ref.~\cite{BDX:2016akw}). LDM flux was computed using GEANT4 via the GEMC 
 interface~\cite{Ung16}. Hall-A BD geometry implemented in GEMC is shown in Figure 
 \ref{fig:GEMC_geometry}.

The simulation procedure is divided into several steps. It starts by sampling muon 
features 
obtained with FLUKA simulations (see Section~\ref{sec:SecondaryMu}). The 
multi-dimensional distribution that includes three-momentum, production vertex, 
statistical weights, and total yield per electron on target (EOT) was converted in the LUND 
format (particle ID, vertex and momentum), and fed to  GEMC.
The interaction of muons with nuclei that produces a new hypothetical dark matter scalar 
particle S  was added to the GEMC process list (details of the theoretical model are 
presented in Section~\ref{sec:DarkScalar}).
The process was implemented according to the prescription described in 
Ref.~\cite{Marsicano:2018vin}, with a more precise production cross section and 
subsequent propagation and decay \cite{Chen:2017awl}. The new class, \texttt{G4Scalar}, 
containing a \texttt{G4ParticleDefinition} instance to include the new $S$ particle was 
implemented in GEMC libraries.
The class initialization requires two parameters: the mass of the scalar and the coupling to 
standard model (SM). This allows one to dynamically set the particle properties at the 
beginning of the simulation. The LDM particle is then set to be unstable, with a lifetime 
that is analytically evaluated following Ref.~\cite{Chen:2017awl}. A single decay channel 
(S $\to \gamma \gamma$) was implemented using the standard GEANT4 
\texttt{G4PhaseSpaceDecayChannel} routine.
In Section~\ref{sec:Scalar10GeV} (Section \ref{sec:Scalar20GeV}), we describe the main 
characteristics of LDM scalar flux produced by the interaction of the high-intensity 
11~GeV 
(22~GeV) electron beam with the Hall-A BD. Finally, the expected sensitivity of a compact 
detector located $\sim$29~m downstream of the BD as a function of the $S$ mass and 
coupling constant is reported.


\section{Secondary Muon Beams}
\label{sec:SecondaryMu}
High-intensity muon beams have applications in many research fields spanning from 
fundamental particle physics~\cite{aguillard2023measurement} to material 
science~\cite{yaouanc2011muon} and inspection and imaging~\cite{instruments6040077}. 
In particular, the use of high-intensity~GeV-energy muon beams could lead to the 
discovery of new light particles not predicted by the SM. \\
Most of the current~\cite{triumf, psi-muon-facilities, MIYAKE200922, ISIS-muon, 
Ganguly:2022ufq, cern-m2-site} and planned~\cite{sns-muon, KIM2020408, 
Chen:2023uvp} facilities produce muons as secondary particles via the decay of 
pions/kaons created by the interaction of an intense proton beam, typically of several MW 
power,  with a heavy material target.  A high-intensity multi-GeV electron beam hitting a 
thick target is likewise a copious source of muons. In this case, muons are produced via 
two classes of processes: 
\begin{itemize}
    \item Photo-production of $\pi$'s and $K$'s, which subsequently decay into muons;
    \item Direct $\mu^+ \mu^-$ pair production.
\end{itemize}

In the latter, muons with energy to the order of the primary electron beam energy are 
produced through a two-step process. First, an electron radiates a $\gamma$ in the 
nucleus field. Secondary particles are then photo-produced nearby. The production 
through a virtual photon exchange (direct electro-production) is instead 
negligible~\cite{COX196977}.
Radiated muons are strongly peaked in the forward direction with energy comparable to the primary beam energy.
Instead, muons produced via decay in flight of photo-produced $\pi$'s and $K$'s show a lower energy spectrum. Monte Carlo simulations of muons produced by the interaction of CEBAF 11~GeV (22~GeV) e$^{-}$-beam with Hall-A BD are shown in  
 Section~\ref{Sect_11GeV_muon} (Section \ref{Sect_20GeV_muon}).   
Distributions are shown for muons with energy grater than 100 MeV.


\subsection{11~GeV Electron Beam}
\label{Sect_11GeV_muon}

To simulate the production and propagation of muons, $\sim 5 \times 10^8$ primary 
electrons with momentum $p_{e^-} = 11$~GeV were generated using the biasing technique 
described in Section~\ref{sect_FLUKA} to reach the target statistics. Results are provided 
per EOT so that they can be easily extrapolated to $10^{22}$ EOT.

\begin{figure}[H]
 \centering
    \includegraphics[scale=0.35]{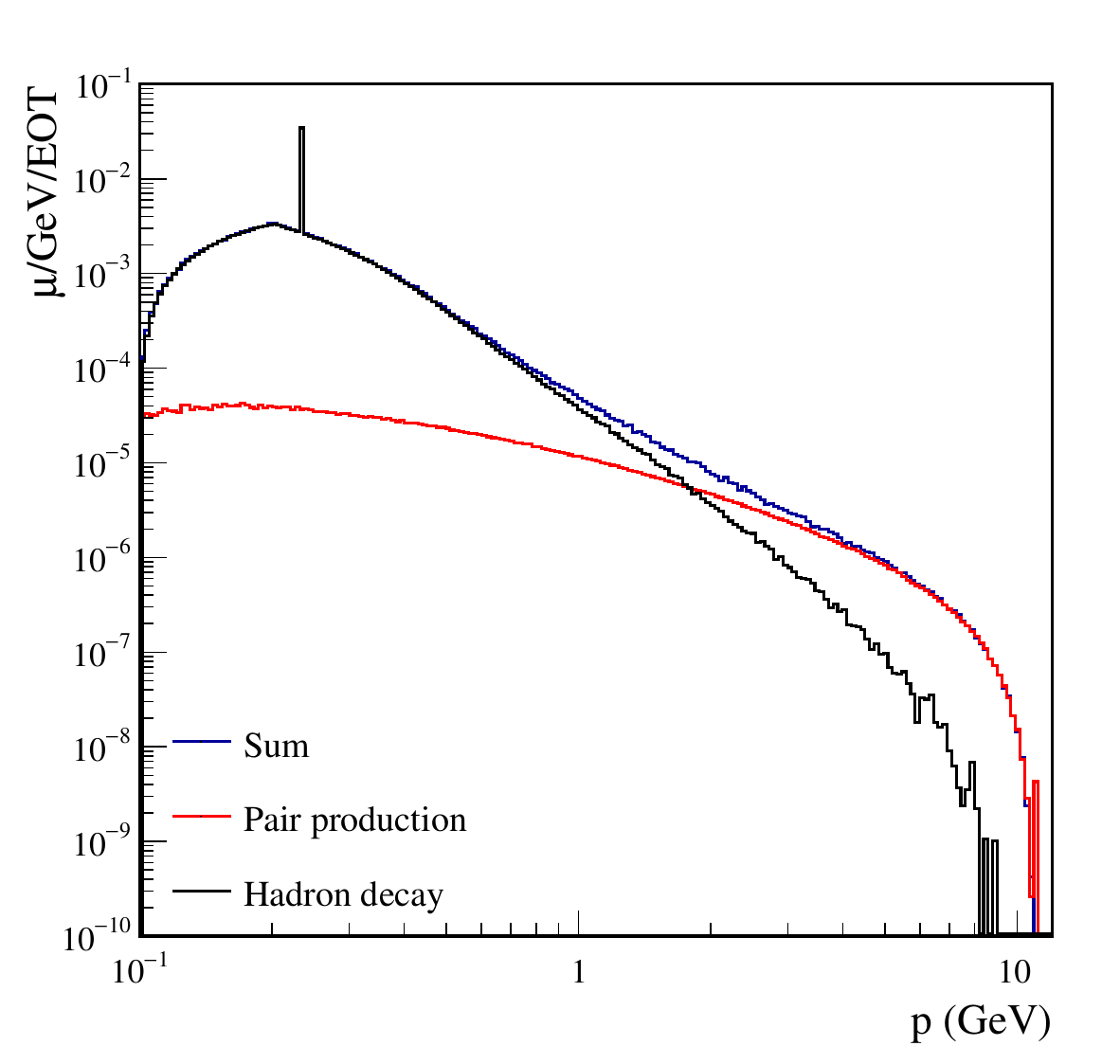}
    \caption{Momentum spectrum of muons at the boundary between the inner core of BD 
    and the outer one with $p>$~100~MeV produced by 11~GeV (blue) electron beam. Pair 
    production is shown in red and hadron decay is shown in black. The ratio between the 
    integrated red and blue spectra is $\sim 15$. The peak at 235~MeV is due to the kaon 
    decay-at-rest process $K\xrightarrow{}\mu+\nu_{\mu}$.}
    \label{fig:beam_comp}
\end{figure}

Momentum distributions of all muons produced in the BD resulting from the two 
production mechanisms (radiation and hadrons decay), are shown in 
Figure~\ref{fig:beam_comp}. The muon flux has been sampled at the boundary between 
the inner core of BD and the outer one. Decay in flight of $\pi$’s and $K$’s dominates 
muon production below 2~GeV, while pair productions dominate at higher energies.

\vspace{-12pt}

\begin{figure}[H]
 \centering
    \includegraphics[scale=0.35]{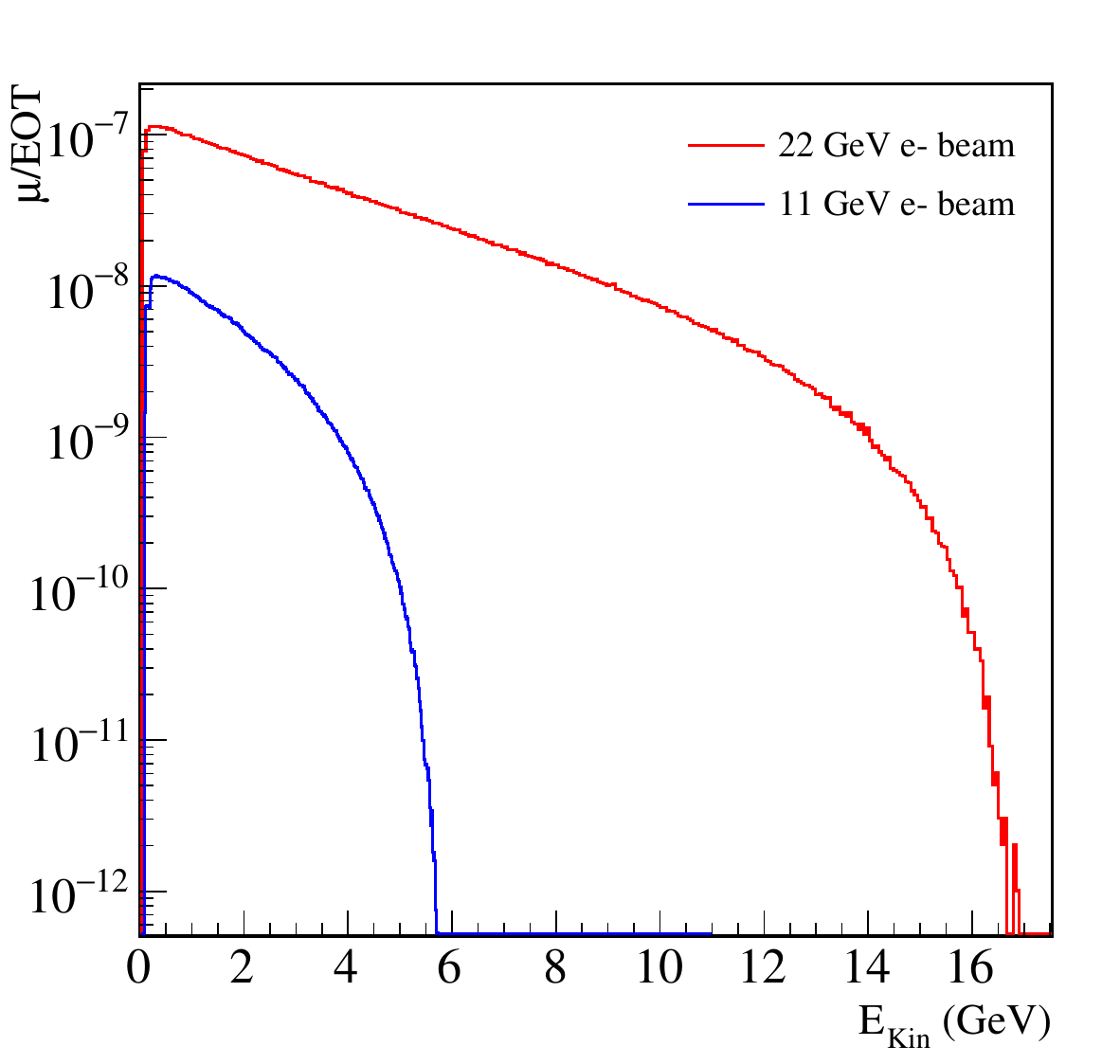}
    \caption{Muon energy distributions produced by an 11~GeV (blue line) and a 22~GeV 
    (red line) CEBAF electron beam interacting with Hall-A BD.}
    \label{fig:Muon_Spectra}
\end{figure}

The kinetic energy distribution of muons produced by the 11~GeV CEBAF  electron beam 
interacting with Hall-A BD is represented by the blue line in 
Figure~\ref{fig:Muon_Spectra}. The flux was computed on a sampling plane (1~m$^{2}$) 
located 10~m downstream the beam dump and perpendicular to the primary e- beam 
direction (corresponding to the light blue thick line in Figure~\ref{fig:geometry_YZ}).
The resulting muon yield per EOT, integrated over $p_\mu>$~100~MeV, is  
$\sim$10$^{-6}$. Therefore, for a primary e$^{-}$-beam current of 50~\textmu A, the 
corresponding muon rate is $\sim$10$^{8}$~$\textmu$/s. These results show the 
advantage of secondary muon beams produced at multi-GeV electron BD facilities in 
comparison to the typical intensity of existing proton beam-produced muon beams with 
similar energies (the Fermilab accelerator complex, for example, can deliver a muon beam 
of about 10$^{7}$~$\textmu$/s at the so-called \textit{magic momentum} of about 3~GeV 
\cite{refId0}).

\vspace{-12pt}

\begin{figure}[H]
 \centering
    \includegraphics[width=0.43\textwidth]{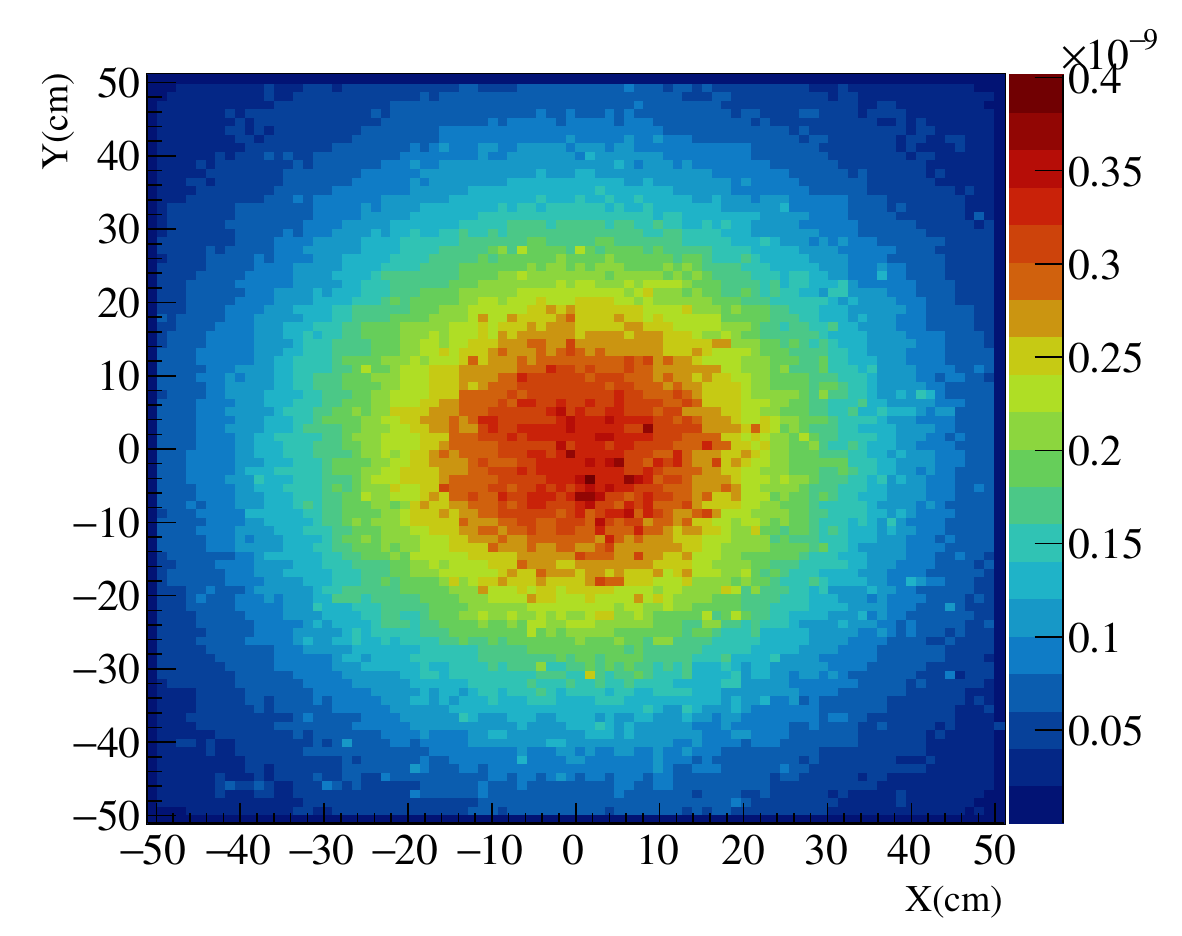}
    \includegraphics[width=0.43\textwidth]{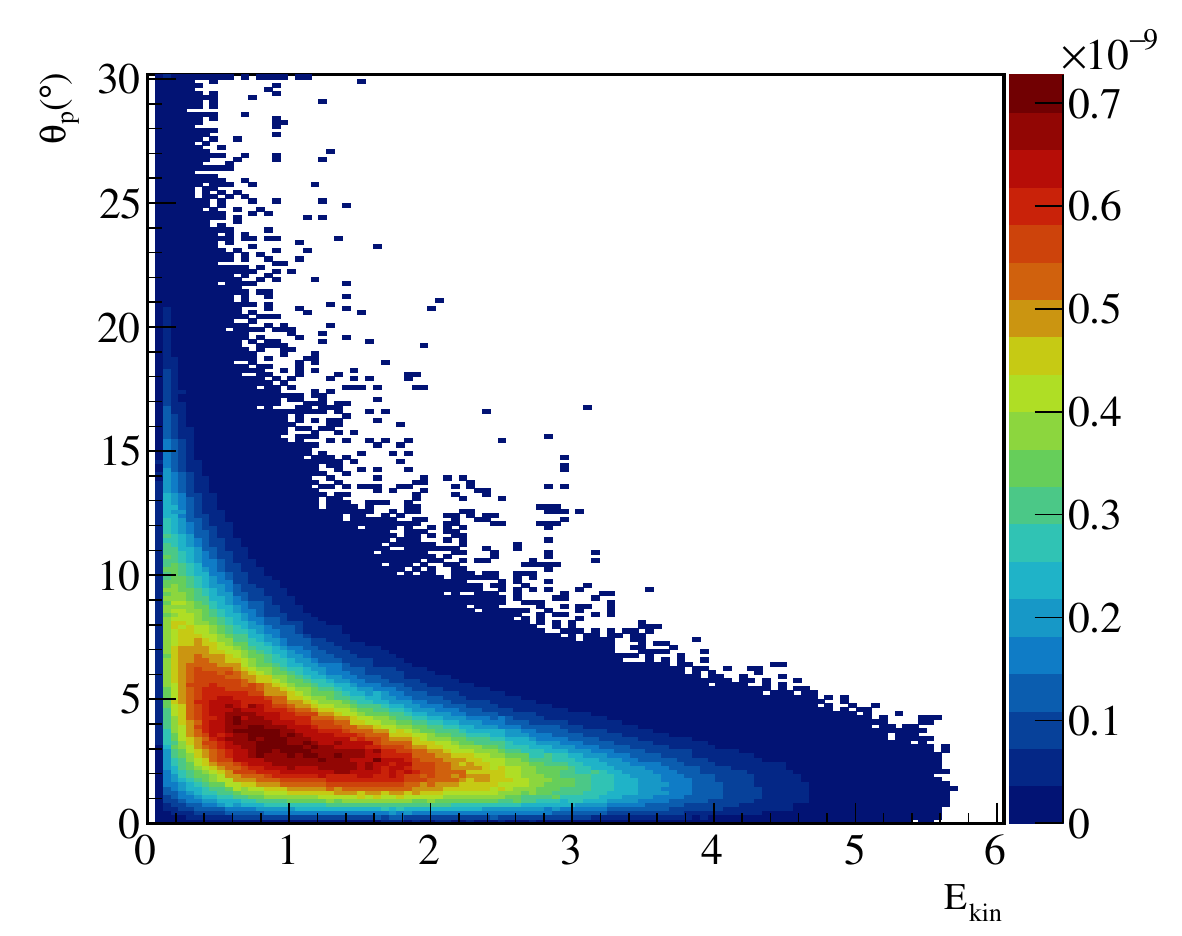} 
    \\
    \includegraphics[width=0.43\textwidth]{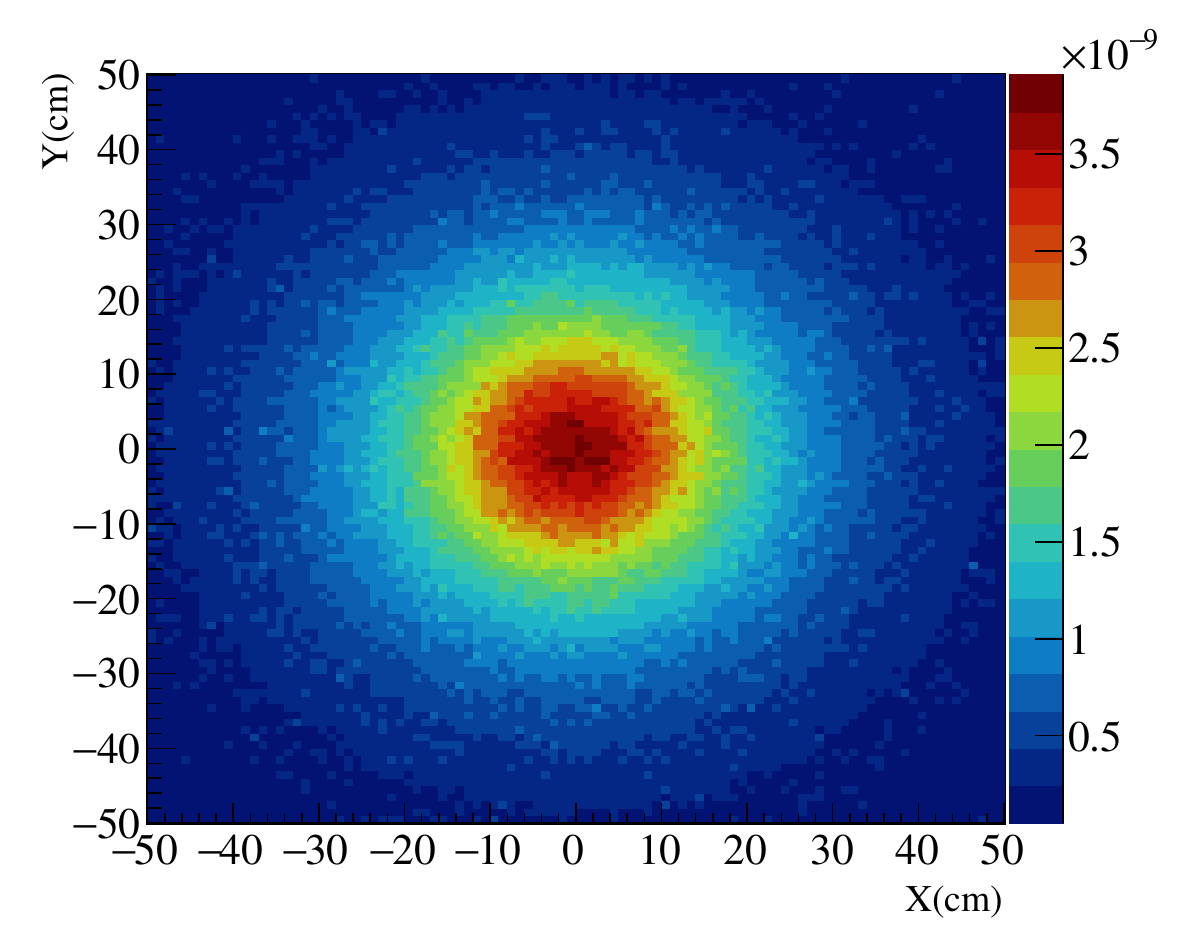}
    \includegraphics[width=0.43\textwidth]{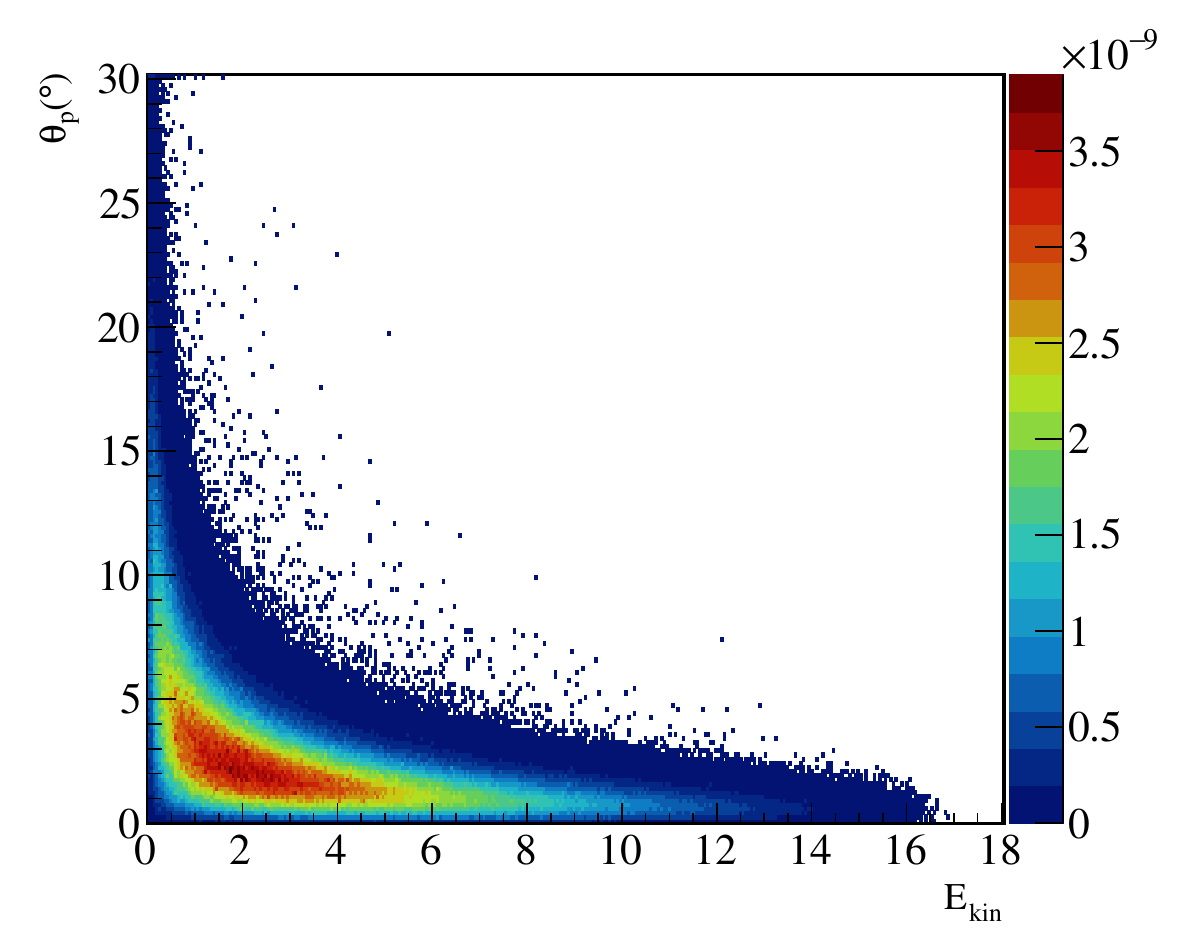}
    \caption{Upper left (lower left): spatial distribution of muons produced in the 
    interaction 
    of CEBAF 11~GeV (22~GeV) electron beam with Hall-A BD. 
    Upper right (lower right): muon angular distribution as a function of energy for 11~GeV 
    (22~GeV) electron beams. }
    \label{fig:Muon_Distr}
\end{figure}

Figure~\ref{fig:Muon_Distr} (top) shows the muon spatial distribution and the direction 
($\theta$ angle) of muons on the sampling surface:  $\sim$50$\%$ of the muons cross the 
plane within an area of roughly $50~\times~50$~cm$^2$. The higher-energy muons are 
mostly produced in the forward direction, while the angular distribution becomes wider 
for lower energies.

\vspace{-3pt}

\subsection{22~GeV Electron Beam}
\label{Sect_20GeV_muon}

A similar simulation of $\sim$$10^{10}$ EOT was performed assuming a CEBAF 22~GeV primary electron beam.

The resulting muon energy distribution is shown in Figure~\ref{fig:Muon_Spectra} with a 
red line. The spectrum remains Bremsstrahlung-like, similar to the 11~GeV case, but it 
covers an extended energy range (up to $\sim$16~GeV) with an almost $\times 8$ yield. 
The spatial distribution (see Figure~\ref{fig:Muon_Distr}, bottom) results in being more 
forward-peaked, with the majority of muons lying on a narrower  
$\sim$40~$\times$~40~cm$^2$ area.

The main characteristics of the muon beams produced by the interactions of 11~GeV and 
22~GeV CEBAF electron beams with the Hall-A BD are summarized in Table 
\ref{tab:muon_summary}.

\begin{table}[H]
 \caption{Summary 
 of JLab secondary muon beam features.}
\label{tab:muon_summary}
\newcolumntype{C}{>{\centering\arraybackslash}X}
\begin{tabularx}{\textwidth}{CCCCC}
\toprule
\multirow{2.5}*{\textbf{Beam Energy}} & \multicolumn{2}{c}{\boldmath\textbf{Flux $\textmu$/EOT}}  & \multirow{2.5}{*}{\boldmath\textbf{$\sigma_x$ (cm)}} & \multirow{2.5}{*}{\boldmath\textbf{$\sigma_y$ (cm)}}  \\
\cmidrule{2-3}
& \boldmath\textbf{100~$\times$~100~cm$^{2}$}  & \boldmath\textbf{25$~\times~$25~cm$^{2}$} & &\\
\midrule
11 GeV& 9.8~$\times~10^{-7}$ & 1.5$~\times~10^{-7}$ & 24.6 & 25.1  \\
\midrule
22 GeV& 7.6~$\times~10^{-6}$ & 1.9~$\times~10^{-6}$ & 20.9 & 20.9 \\
\bottomrule
\end{tabularx}%

\end{table}

\section{Secondary Neutrino Beams}
\label{sec:secondaryNu}
Fission reactors and proton accelerators are currently the main source of neutrino beams. 
The reactors produce electron-type antineutrinos from fission fragment beta decay and are 
widely used in low-energy ($\sim$MeV)  experiments. In accelerators, high-energy 
protons hit a target to generate short-lived hadrons (mainly $\pi^{\pm}$ and $K^{\pm}$) 
that successively either decay in flight (DIF) or decay at rest (DAR) into neutrinos. \\
DAR neutrinos, mainly produced by spallation neutron sources~\cite{PhysRevD.106.032003}, show an isotropic spatial distribution with an energy spectrum depending on the decay:
\begin{itemize}
    \item $\pi^{+}\xrightarrow{}\mu^{+}+\nu_{\mu}$,  E$_\nu \sim$ 29.8~MeV, almost monochromatic;
    \item  $\mu^{+}\xrightarrow{}\bar\nu_{\mu}+\nu_{e}+e^{+}$, E$_\nu$ in the range 0--52.8~MeV;
        \item  
        $K^{+}\xrightarrow{}\mu^{+}+\nu_{\mu}$, E$_\nu \sim$ 236~MeV, almost monochromatic.
\end{itemize}

DAR neutrinos are suitable for studying coherent elastic neutrino-nucleus scattering 
(CEvNS). This process, predicted a long time ago, has been only recently 
observed~\cite{PhysRevLett.126.012002} and is a leading candidate for the study of 
non-standard (BSM) neutrino interactions~\cite{PhysRevD.107.055019}.

Proton beam dump facilities are also used as high-intensity secondary neutrino beam generators~\cite{ATHANASSOPOULOS1997149, PhysRevD.79.072002}. 

In the following two sections, we will report the results of simulations of the 11~GeV 
(22~GeV) CEBAF electron beam interaction with the Hall-A BD,
showing that this can be used as an  efficient and high-intensity source of DAR neutrinos.

\subsection{11~GeV Electron Beam}

\begin{figure}[H]
 
\centering
\includegraphics[width=0.8\columnwidth]{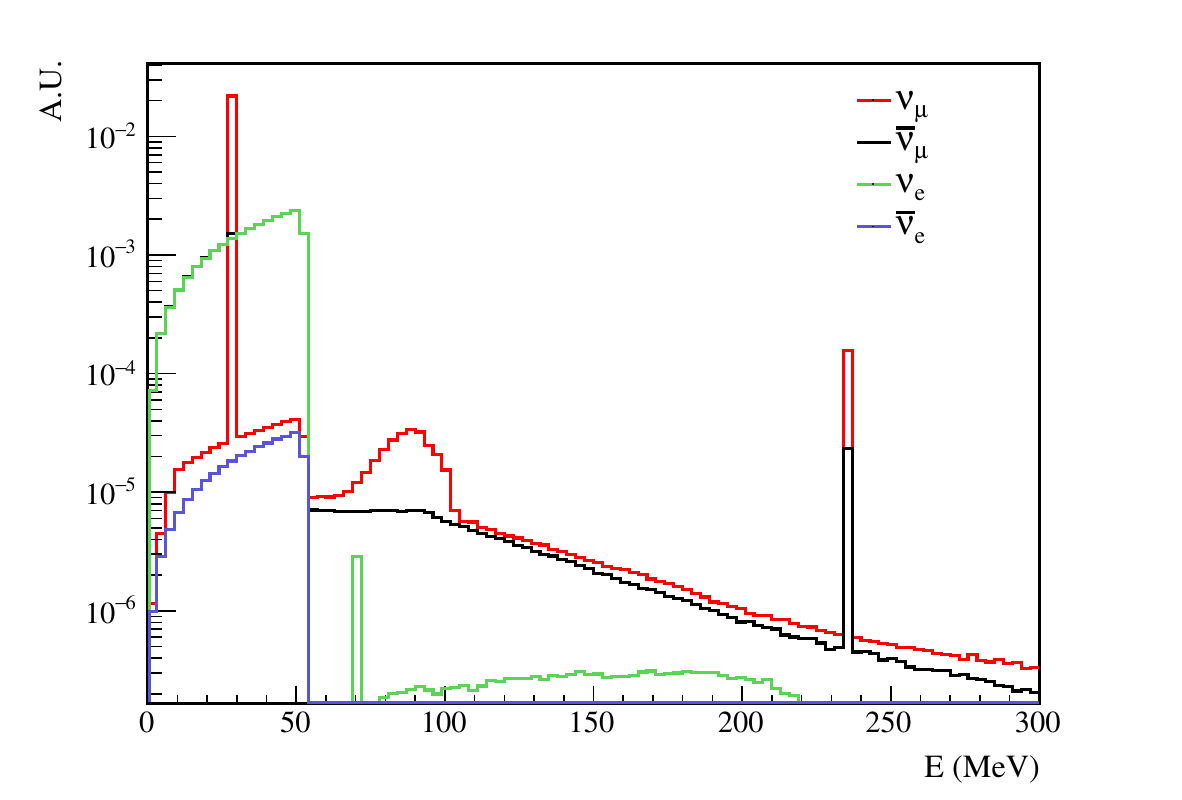}
\caption{Neutrino energy spectrum produced by the interaction of the CEBAF 11~GeV e$^-$ beam with the Hall-A BD. Each color corresponds to a different neutrino species, as the legend reports.}
\label{neutrino_flux_family}
\end{figure}

To simulate the production and propagation of neutrinos produced by the interaction of the CEBAF 11~GeV electron beam with the Hall-A BD, the procedure described in Section~\ref{sect_FLUKA} was used.
Figure~\ref{neutrino_flux_family} shows the resulting neutrino energy spectrum.
As anticipated in the previous section, a peak around 29.8~MeV and another peak 
236~MeV related to $\pi$ and $k$ DAR, are clearly visible over a smooth background due 
to the muon decay and DIF events. The peak at 70~MeV has been tracked back to pion 
decay in electron and electronic neutrino. As expected, it is suppressed by four orders of 
magnitude with respect to the dominant allowed decay 
$\pi^{+}\xrightarrow{}\mu^{+}+\nu_{\mu}$.

\vspace{-12pt}

\begin{figure}[H]
 \centering
    \includegraphics[width=0.43\textwidth]{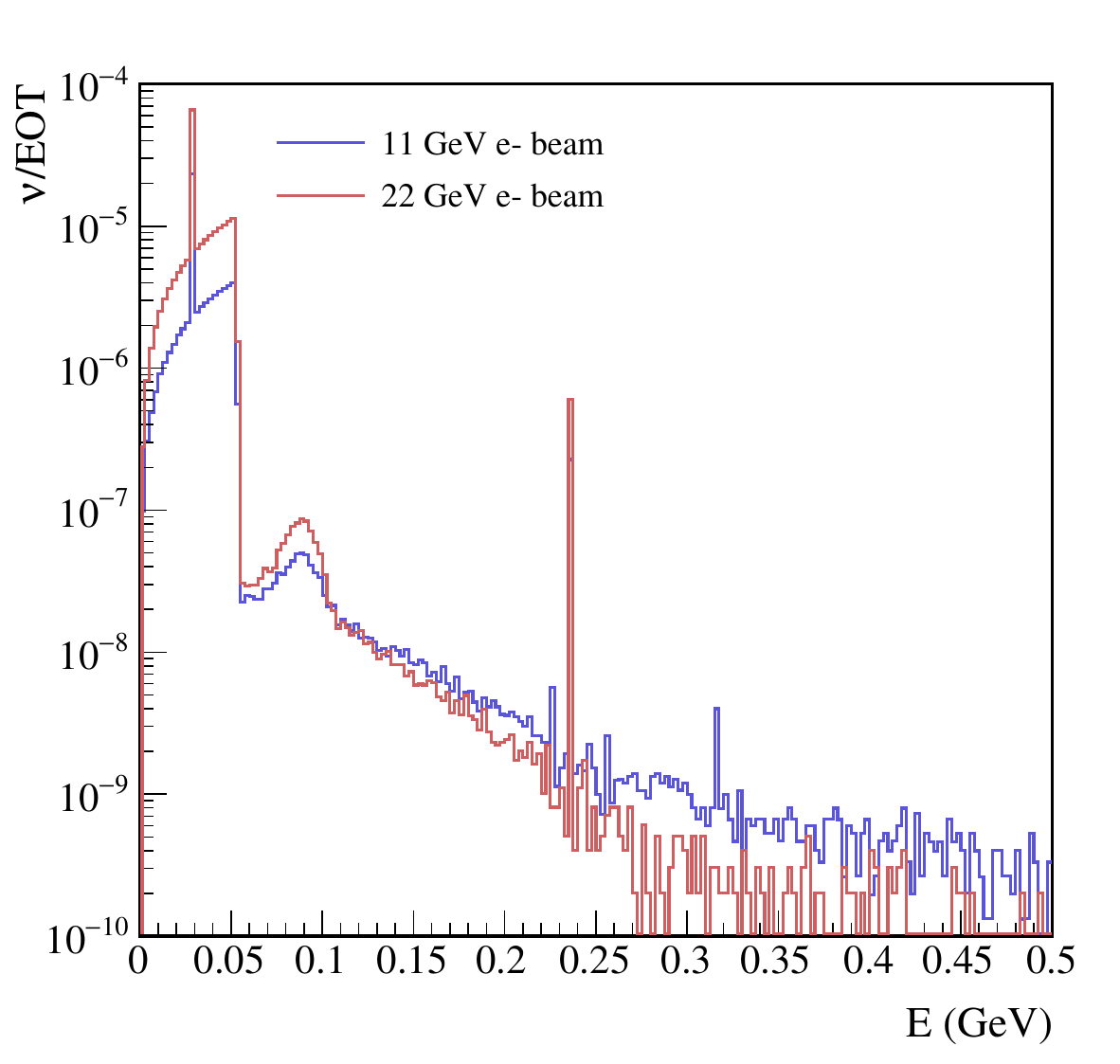}
    \includegraphics[width=0.43\textwidth]{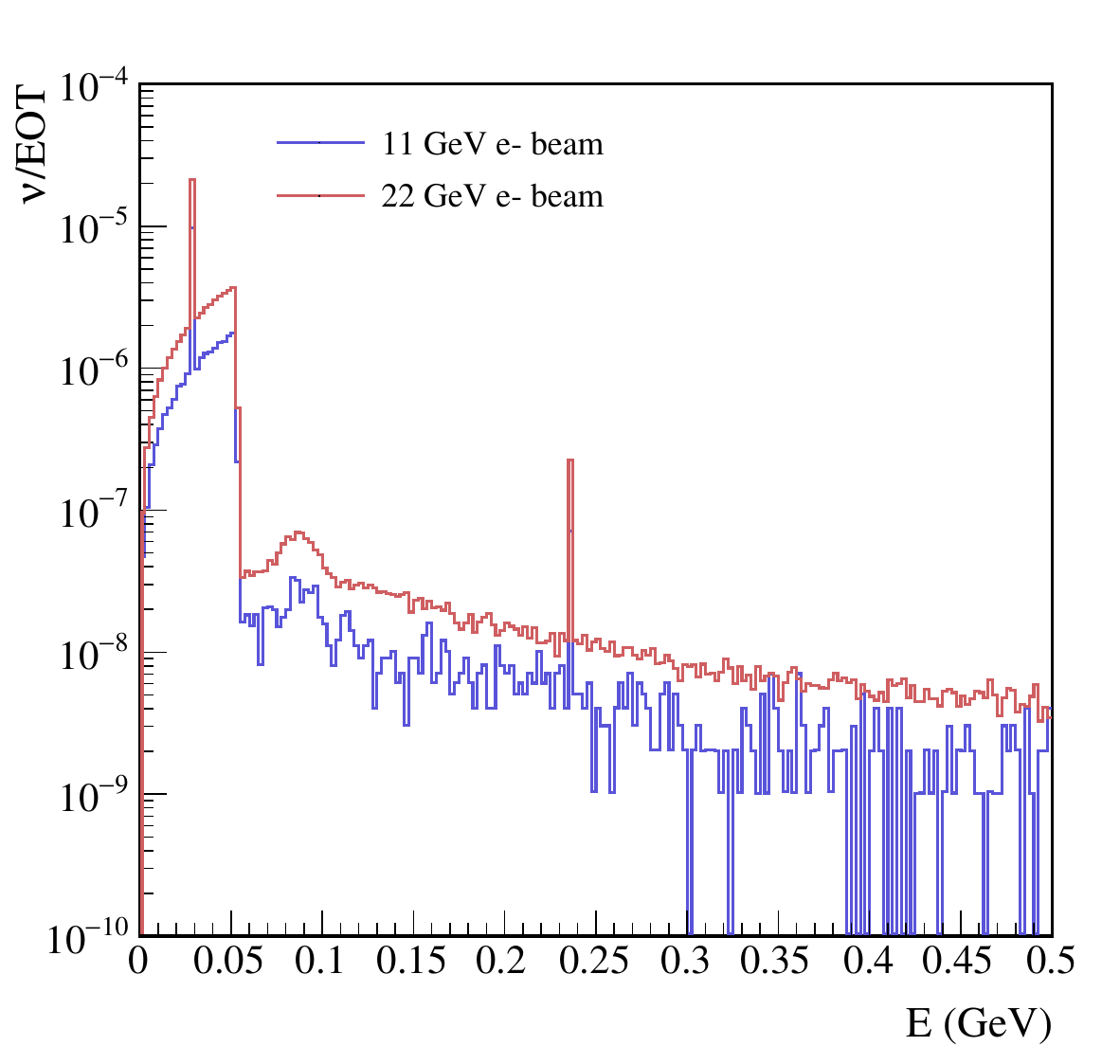}
    \caption{Energy distribution of off-axis (left panel) and on-axis (right panel) neutrinos produced by the interaction of 11~GeV (blue) and 22~GeV (red) CEBAF electron beams with the Hall-A BD.}
    \label{fig:neutrino_11GeVvs20GeV}
\end{figure}

We studied the characteristics of the neutrino flux produced along the primary electron beam direction (on-axis) and perpendicular to it (off-axis). 
For the latter, we computed the flux on a  1~m$^{2}$ sampling plane located $\sim$10~m 
above the dump corresponding to the ground level (orange surface in 
Figure~\ref{fig:geometry_YZ}). 
Results show that the off-axis $\nu$ energy spectrum (see 
Figure~\ref{fig:neutrino_11GeVvs20GeV}, left panel) is compatible with the spectrum of a  
DAR source. The overall neutrino flux in the energy range 0--100~MeV  is $\sim 6.6 \times 
10^{-5}$ $\nu$/EOT, corresponding to 99$\%$ of the spectrum. Therefore, for an 
accumulated charge of 10$^{22}$ EOT per year,  an intense flux of $\sim10^{18}~\nu$, 
comparable to the integrated flux of the flagship DAR-neutrino facility SNS@Oak Ridge 
National Lab~\cite{PhysRevD.106.032003}, is expected. 

\vspace{-6pt}

Figure~\ref{fig:neutrino_11GeVvs20GeV} (right panel) shows the energy distribution of 
on-axis neutrinos. 
The neutrino flux was sampled on a 1~m$^2$-plane downstream of the BD at the exit 
surface  of the concrete shielding (the same used for sampling muons in 
Section~\ref{sec:SecondaryMu}, corresponding to the light blue thick line in 
Figure~\ref{fig:geometry_YZ}). Even if the DAR contribution is dominant, a tiny but 
non-negligible part of the spectrum shows energies greater than 100~MeV. 
The resulting on-axis neutrino flux in the energy range 0--500~MeV is $\sim$$2.9\times 10^{-5}$~$\nu$/EOT, with the DAR part corresponding to $\sim$96$\%$ of the overall yield.  

\subsection{22~GeV Electron Beam}
Similar to the previous paragraph, we evaluated the neutrino flux produced by the 
interaction of a primary 22~GeV e$^{-}$-beam with  Hall-A BD. 
Figure~\ref{fig:neutrino_11GeVvs20GeV} compares the on- and off-axes neutrino energy 
distributions produced by an 11~GeV and 22~GeV electron beam. They show a similar 
shape, with a yield difference of about a factor of two. More precisely, the results of 
simulations show an overall off-axis (on-axis) flux of  $\sim$$1.9 \times 10^{-4}$ 
($6.3\times 10^{-5}$) neutrino/EOT in the energy range 0-500~MeV.

In Table \ref{tab:neutrino_summary} the characteristics of neutrino fluxes are summarized. 

\begin{table}[H]
 \caption{Summary of JLab secondary neutrino beam features. Yields are obtained integrating the neutrino flux in the energy range 0--500~MeV.}
\label{tab:neutrino_summary}
\newcolumntype{C}{>{\centering\arraybackslash}X}
\begin{tabularx}{\textwidth}{CCC}
\toprule
\boldmath\textbf{Beam Energy}  & \boldmath\textbf{Off-Axis Flux [$\nu$/EOT/m$^2$]}  & \boldmath\textbf{On-Axis Flux [$\nu$/EOT/m$^2$] } \\
\midrule
11 GeV& $6.7 \times 10^{-5}$ & $2.9 \times 10^{-5}$  \\
\midrule
22 GeV & $1.9 \times 10^{-4}$ & $6.3 \times 10^{-5}$  \\
\bottomrule
\end{tabularx}%

\end{table}

\section{Dark Matter Beams}
\label{sec:DarkScalar}
Despite several years of research, the particle nature of dark matter remains one of the 
biggest endeavors in fundamental science  (for a review, 
see~\cite{RevModPhys.90.045002}). Huge efforts have been spent in recent years into its 
identification, focusing on the search of weakly interacting massive particle candidates 
(WIMPs) with masses in the range 1~GeV--10~TeV. The lack of experimental evidences has 
motivated the interest toward sub-GeV light dark matter (LDM)  where direct detection 
has a limited sensitivity ~\cite{Battaglieri:2017aum, Krnjaic:2015mbs, 
Fabbrichesi:2020wbt,Filippi:2020kii}.
To achieve the correct abundance inferred from astrophysical constraints, the interaction 
between LDM and SM states has to be mediated by a new gauge group that is light force 
carrier neutral under the standard model. 
The existence of the LDM would also bring theoretical predictions in agreement with 
observations~\cite{Liddle:1998ew,Coy:2021ann} such as reconciling the persistent 
$\sim$4$\sigma$ discrepancy in the anomalous magnetic moment of the 
muon~\cite{Muong-2:2004fok, Muong-2:2021ojo}.

The theoretical options for the inclusion of  new interactions and particles in the SM are  
limited. A minimal list includes: 
 the renormalizable vector portals mediated by a dark vector
boson; scalar portals mediated by a new scalar mixing
with the Higgs boson; and neutrino portal operators,
mediated by a heavy neutral lepton. Vector  and scalar portals
are particularly motivated for a  dark matter with mass 
in the  MeV-GeV range. In parallel, a significant experimental activity has been performed 
to verify or falsify the different hypothesis. Data collected in previous experiments, 
optimized for different physics scopes, were re-analyzed within the above-mentioned  
theoretical frameworks, providing exclusion limits in the parameter space of the theories. 
New experiments, specifically designed to investigate the different options, 
have already collected data or are expected to run in the near future.
We refer the interested reader to the {\it Feebly-interacting particles: FIPs 2022 Workshop Report} ~\cite{Antel:2023hkf} for a comprehensive discussion about the possible LDM theoretical scenarios, the current experimental efforts, and a survey on the future proposals to detect LDM. 

In this work, we focused on a minimal model that could explain the $(g-2)_\mu$ anomaly: 
a new \textit{leptophilic} scalar dark matter state (\textit{dark scalar} or $S$) that couples 
only to muons. A detailed description of the theoretical model is reported in 
Ref.~\cite{Chen:2017awl, Marsicano:2018vin} and the references therein. 
In this model, the main process responsible for  $S$ emission by an impinging muon on a fixed target is the so-called “radiative” production $\mu$~+~N$\xrightarrow{}$$\mu$~+~N~+~$S$.
The incident muon interacts with a target nucleus, N, by exchanging a photon, $\gamma$, and radiates the $S$.
 
For the mass range  ($m_S<2m_\mu$), $S$ could only decay  into two photons with a 
decay width, $\Gamma_{\gamma \gamma}$, which depends on the $\mu$--$S$ 
coupling constant, $g_\mu$, and the ratio of muon to $S$ masses, $m_\mu/m_S$ 
~\cite{Chen:2017awl}:
\begin{equation}
    \label{eq:scalar_decay_width}
    \Gamma_{\gamma\gamma} = \frac{\alpha^2m^3_S}{128\pi^3} 
    \left| \frac{g_\mu}{m_\mu} \frac{4m^2_\mu}{m^2_S}\left[ 1+ \left(1-\frac{4m^2_\mu}{m^2_S} \right) \arcsin^2{\left(\frac{4m^2_\mu}{m^2_S}\right)^{-1/2}} \right]\right|^2
\end{equation}

Different experimental techniques can be used to search for muon-coupling light dark scalars. Among them, medium-energy electron beam dump experiments, providing an intense source of secondary muons, cover a broad area in the $g_\mu$ vs.  $m_S$ parameter space, as shown in Ref.  \cite{Marsicano:2018vin}.
As shown in Sections~\ref{Sect_11GeV_muon} and ~\ref{Sect_20GeV_muon}, muons are copiously produced by the interaction of the CEBAF electron beam with the Hall-A BD. They penetrate deeply into the dump and surrounding materials, losing energy mainly through ionization and, while traveling, may radiate a $S$ particle. 

In Sections~\ref{sec:Scalar10GeV} and \ref{sec:Scalar20GeV}, we present the 
characteristics of a hypothetical dark scalar $S$ beam produced respectively by the 
interaction of a primary 11~GeV and 22~GeV electron beam with the Hall-A BD. For the 
former, a realistic evaluation of the background based on data collected at 10~GeV electron 
beam in the BDX-MINI experiment~\cite{Battaglieri:2020lds} was possible. This provided 
a solid ground to realistically evaluate the expected sensitivity of an experiment 
($s$BDX-MINI), which uses a reduced version of the BDX detector~\cite{BDX:2016akw}. 
Results are reported in Section~\ref{sec:discovery}. In the 22~GeV electron beam case, we 
did not evaluate the experimental sensitivity since a realistic background model was not 
available.

\subsection{11~GeV Electron Beam}
\label{sec:Scalar10GeV}

In order to characterize the hypothetical dark scalar beam, $\sim 10^9-10^{11}$ muons were simulated using the biasing procedure described in Section~\ref{sec:GEMC}. Simulations were performed assuming a  fixed coupling constant $g_\mu = 3.87 \times 10^{-4}$ and $m_S$ in the range 25~MeV--210~MeV. To keep the computational time reasonable, a further bias factor of $10^7$ was introduced in FLUKA simulations.

\begin{figure}[H]
 \centering
    \includegraphics[width=0.43\textwidth]{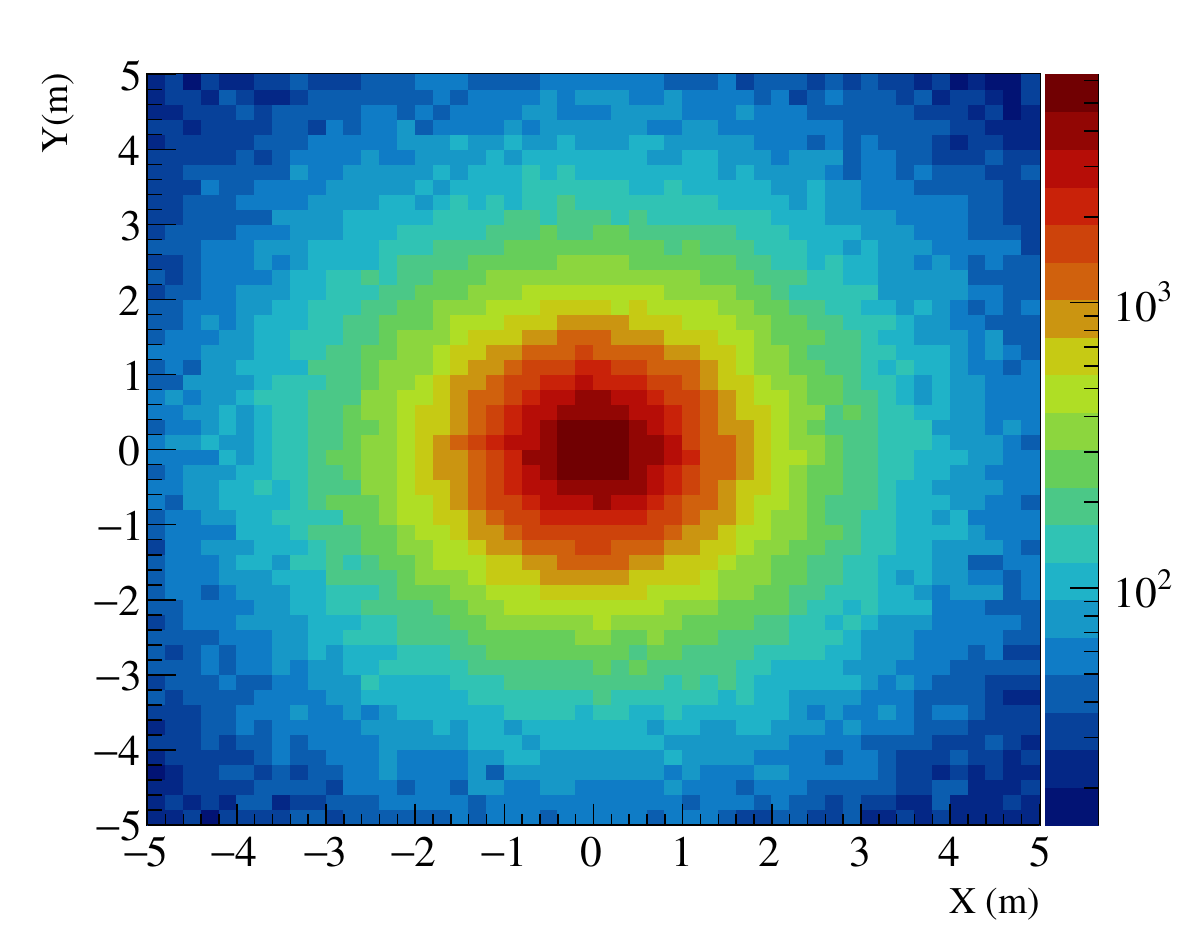}
    \includegraphics[width=0.43\textwidth]{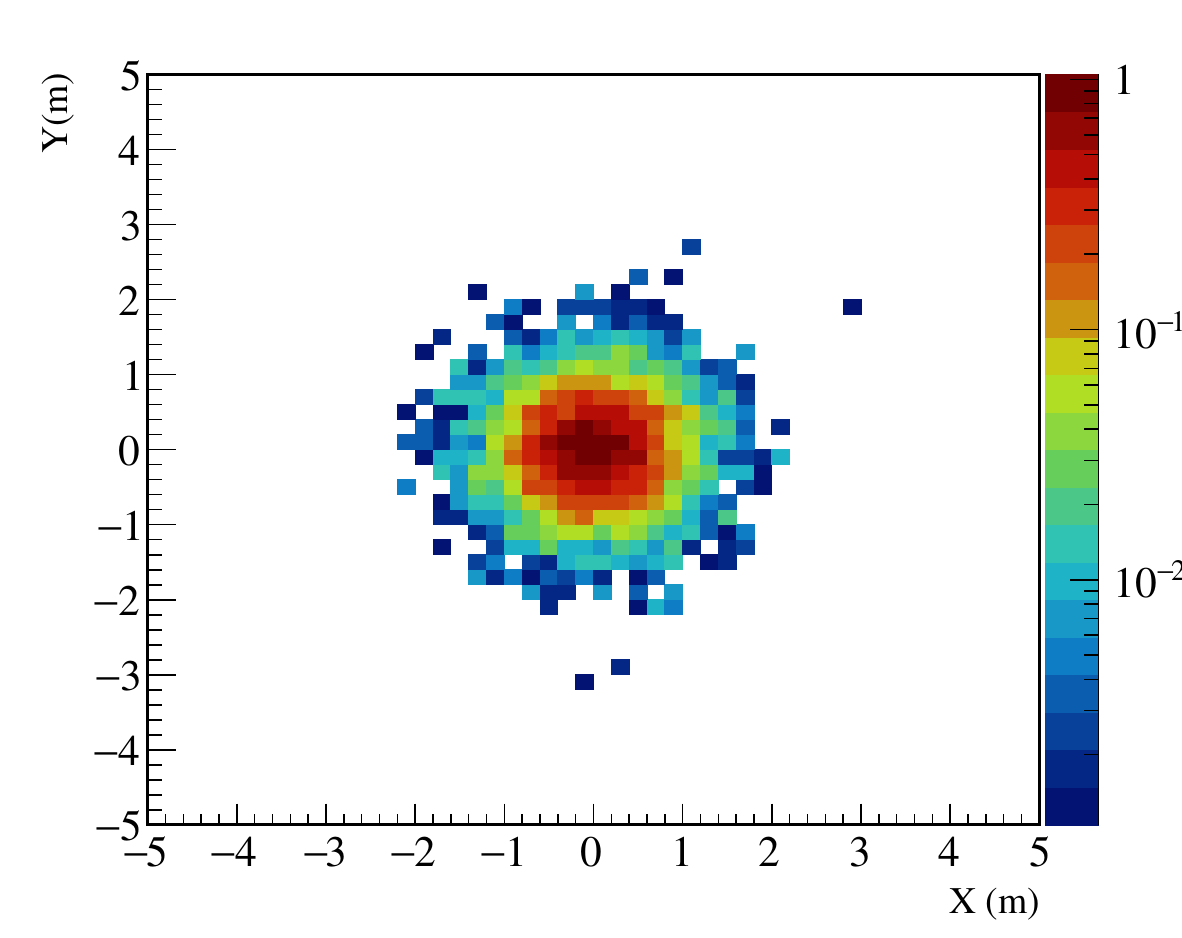}
    \vfill
    \includegraphics[width=0.43\textwidth]{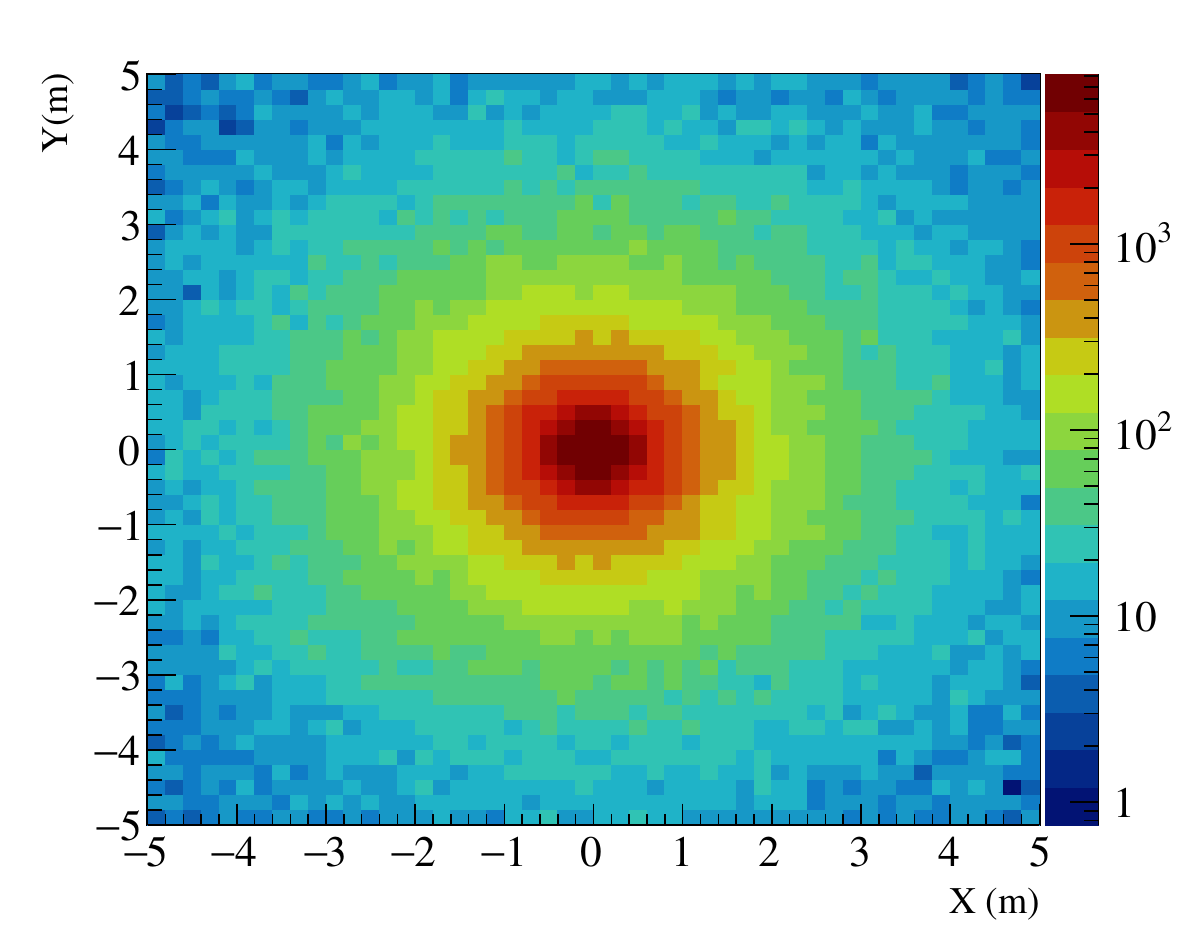}
    \includegraphics[width=0.43\textwidth]{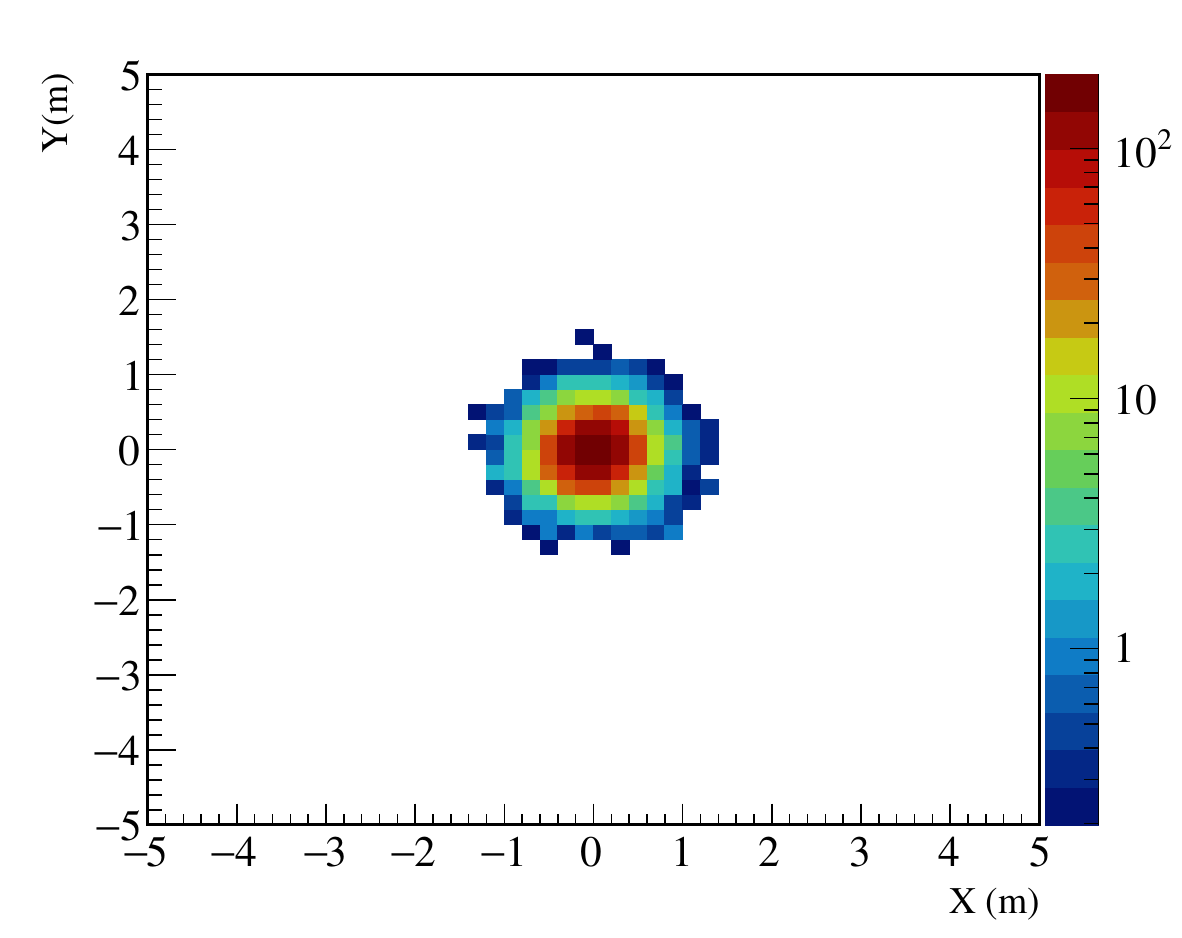}
    
    \caption{Spatial
 distributions of $S$ sampled 20~m downstream of the beam dump. 
    The top (bottom) row refers to an $S$ beam generated by the 11~GeV (22~GeV) CEBAF electron beam. The beam spot size refers to  $m_S$~=~50~MeV (left) and $m_S$~=~180 MeV    (right).}
    
\label{fig:scalar_XY_distrib}

\end{figure}

Figure~\ref{fig:scalar_XY_distrib} 
shows results for the dark scalar beam obtained with an 11~GeV primary electron beam. 
The top panel shows the $S$ spatial distribution on a sampling plane located 20~m 
downstream of the beam dump. The plot on the left was obtained assuming a dark scalar 
mass of $m_S$~=~50~MeV, while the plot on the right refers to $m_S$~=~180~MeV.
The difference in the $S$ beam spot size is due to a different fraction of energy transferred 
from the muon to the radiated $S$ that increases for larger $m_S$ (more energetic $S$ 
corresponds to a smaller spatial spread).

\begin{figure}[H]
 \centering
\includegraphics[width=0.43\textwidth]{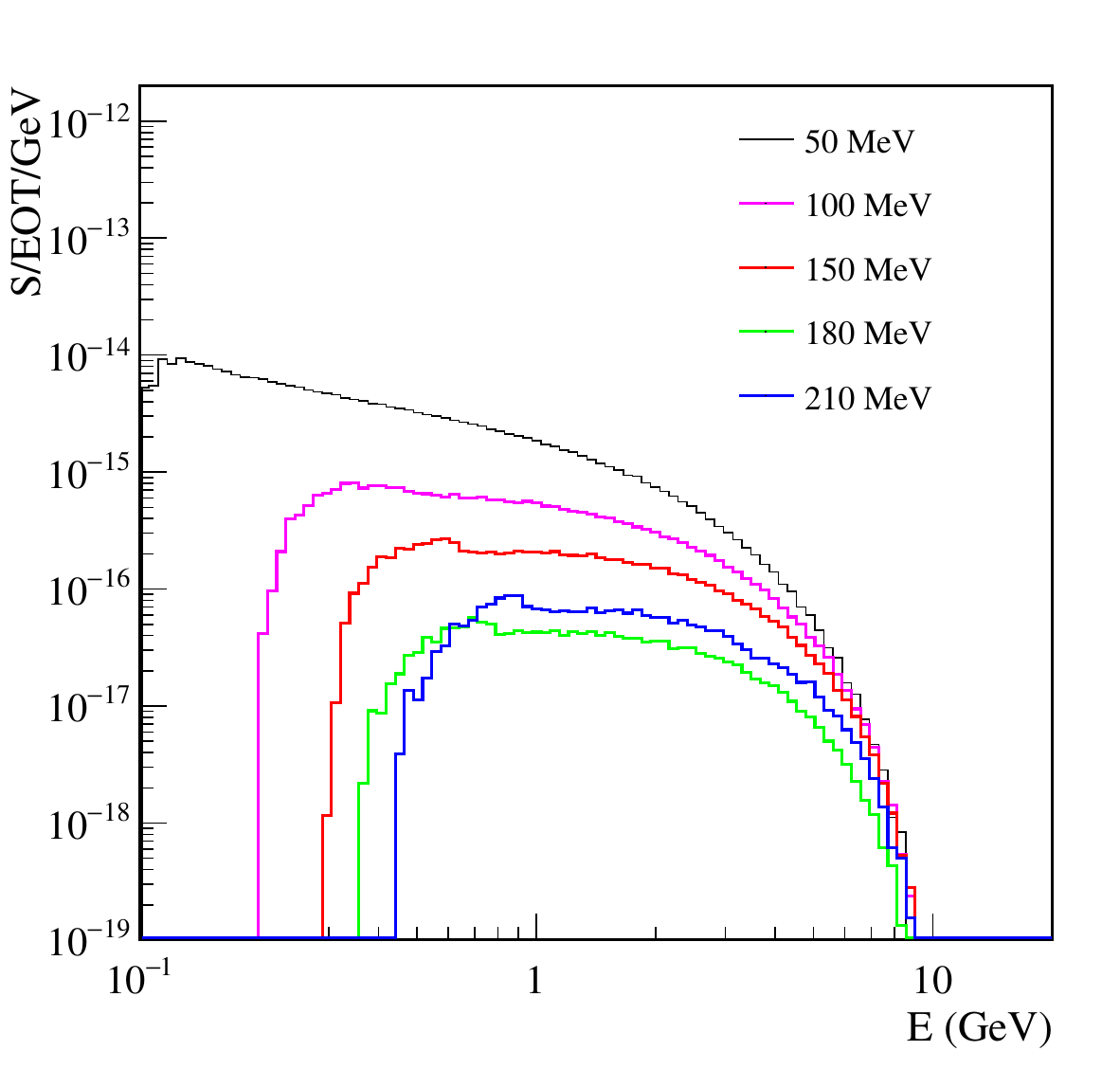}
\includegraphics[width=0.43\textwidth]{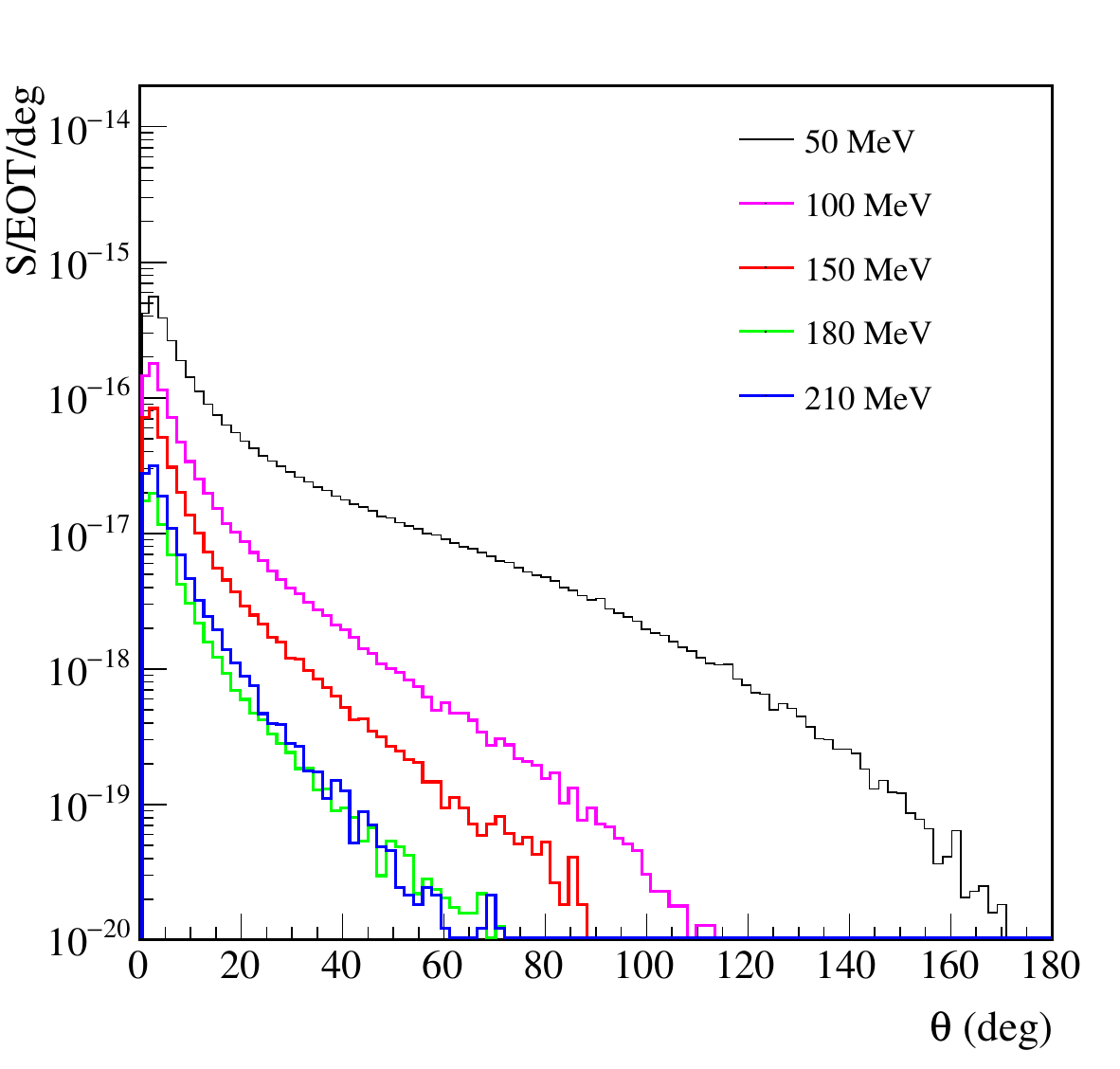}
\vfill
\includegraphics[width=0.43\textwidth]{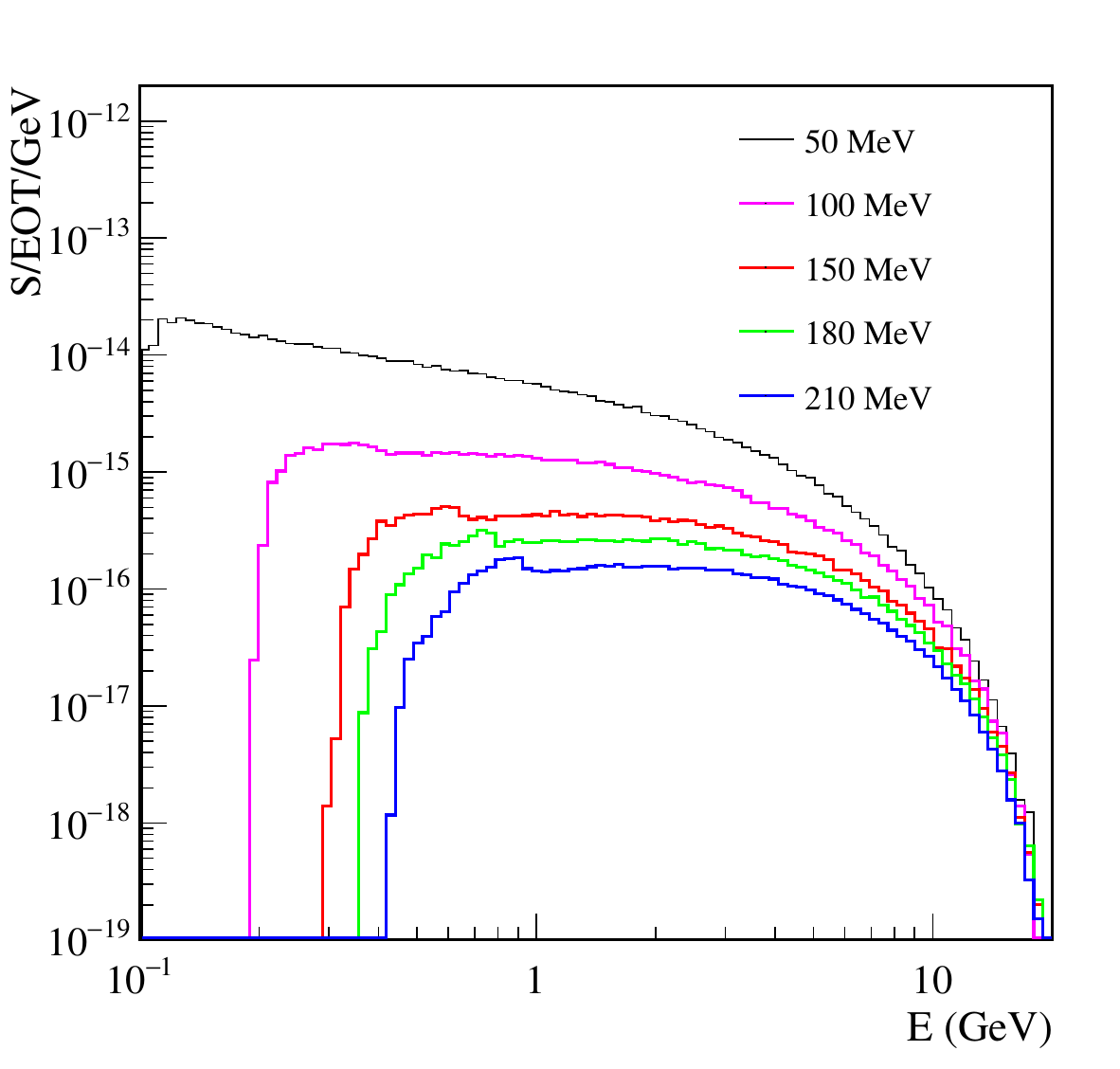}
\includegraphics[width=0.43\textwidth]{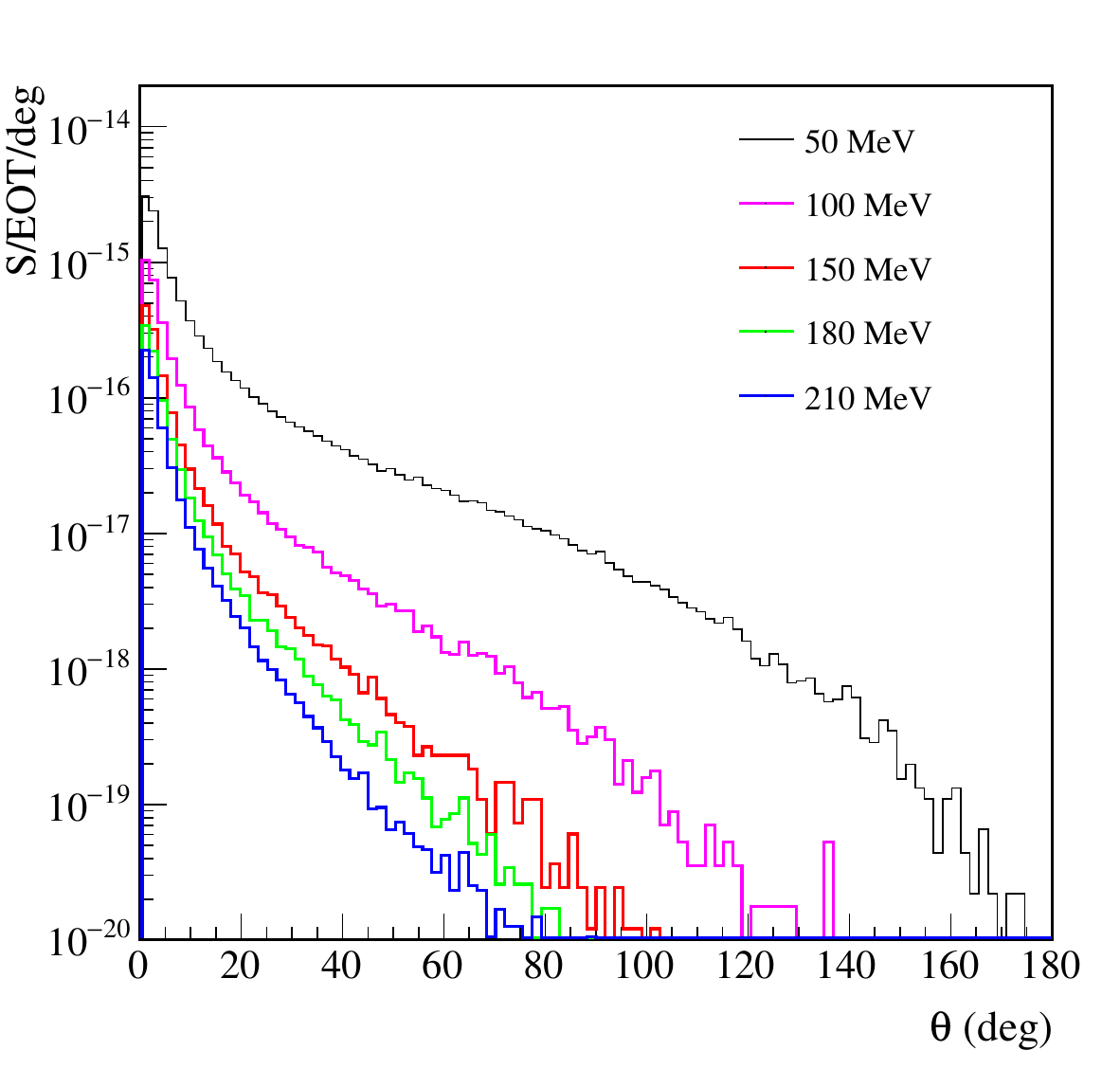}
\caption{Energy (left) and angular (right) distributions of the dark scalar $S$ for different masses. Results are shown for a primary 11~GeV electron beam (top) and 22~GeV (bottom).
\label{fig:scalar_flux}}
\end{figure}

The $S$ energy spectrum is shown, for different $m_S$, on the top-left panel of Figure~\ref{fig:scalar_flux}. The right column shows the $S$ angular distribution with respect to the primary beam direction. All distributions are normalized to the number of $S$ per EOT. The energy distribution for light scalar shows a peak at low energy, since for heavier scalar the out-going $S$ takes a larger fraction of the muon energy. 
The kinematic of the produced $S$ strongly depends on the mass: heavy $S$ are mostly 
produced in the forward direction, while the angular distribution is wider for lighter $S$.

\vspace{-6pt}

\vspace{-15pt}

\subsection{22~GeV Electron Beam}
\label{sec:Scalar20GeV}
Simulations were performed using the same bias factor and coupling $g_\mu$ used for the 11~GeV electron beam case. The resulting beam spot size, energy, and angular distributions are shown in the bottom panels of  Figures~\ref{fig:scalar_XY_distrib} and ~\ref{fig:scalar_flux}, respectively. They show a   behavior similar to the 11~GeV case, with a more focused dark beam spot that covers an extended energy range. The $S$ yield increases by a factor of 3--10, depending on the scalar mass.

Finally, Table~\ref{tab:dark_scalar_beam_sumamry} summarizes the expected  $S$ yield per EOT and beam spot size, sampled in a plane located 20~m downstream of the beam dump for an 11 and 22~GeV beam, and the two values of $m_S$.

\begin{table}[H]
 \caption{Summary
 of JLab scalar dark matter beam features.}
\label{tab:dark_scalar_beam_sumamry}
\newcolumntype{C}{>{\centering\arraybackslash}X}
\begin{tabularx}{\textwidth}{CCCCC}
\toprule
\multirow{2.5}{*}{\boldmath\textbf{Beam Energy (GeV)}} & \multicolumn{2}{l}{\boldmath\textbf{$\;\;\;\;\;\;\;\;\;\;\;\;\; m_S$ = 50 MeV}}                                                              & \multicolumn{2}{l}{\boldmath\textbf{\;\;\;\;\;\;\;\;\;\;\; $m_S$ = 180 MeV}}                                                             \\ \cmidrule{2-5} 
                              & \multicolumn{1}{l}{\begin{tabular}[c]{@{}l@{}}\boldmath\textbf{$S$/EOT}\end{tabular}} & \boldmath\textbf{$\sigma$ (m)} & \multicolumn{1}{l}{\begin{tabular}[c]{@{}l@{}}\boldmath\textbf{$S$/EOT}\end{tabular}} & \boldmath\textbf{$\sigma$ (m)} \\ \midrule
11                            & \multicolumn{1}{l}{$5.27\times10^{-15}$}                                         & 1.556& \multicolumn{1}{l}{$1.32\times10^{-16}$}                                         & 0.488\\ \midrule
22                            & \multicolumn{1}{l}{$1.90\times10^{-14}$}                                         & 1.22& \multicolumn{1}{l}{$1.44\times10^{-15}$}                                   & 0.304\\ \bottomrule
\end{tabularx}

\end{table}

\subsection{Discovery Potential of $s$BDX-MINI Experiment}
\label{sec:discovery}

The two pipes already installed downstream of Hall-A BD could host a new experiment searching for the dark scalar particle $S$: $s$BDX-MINI. The same infrastructure was used for the BDX-MINI experiment~\cite{Battaglieri:2020lds}. In this section, we explored the sensitivity of a BDX-MINI-like experiment searching for $S$ in the visible decay mode ($S \to \gamma \gamma$) with both gammas detected.
The sBDX-MINI would make use of CEBAF 11~GeV e$^{-}$ beam hitting the Hall-A BD 
running for about 1~year, with currents up to 75~\textmu A (corresponding to an 
accumulated charge of 10$^{22}$~EOT). 
We assumed a detector with a layout similar to BDX-MINI, with an almost cylindrical 
electromagnetic calorimeter with an 8~cm radius surrounded by a multi-layer veto system.
To compensate for the limited pipe size (10$^\prime$$^\prime$), we assumed a 2~m 
vertical length detector, roughly corresponding to 4 BDX-MINI detectors stacked. If we 
assume the current JLab setup, some  muons produced by the 11 GeV beam interaction 
with the BD will reach the two pipes. To reduce this background, the detector was 
assumed to be located in the farthest well. 

\begin{figure}[H]
\centering
\includegraphics[width=9.5 cm]{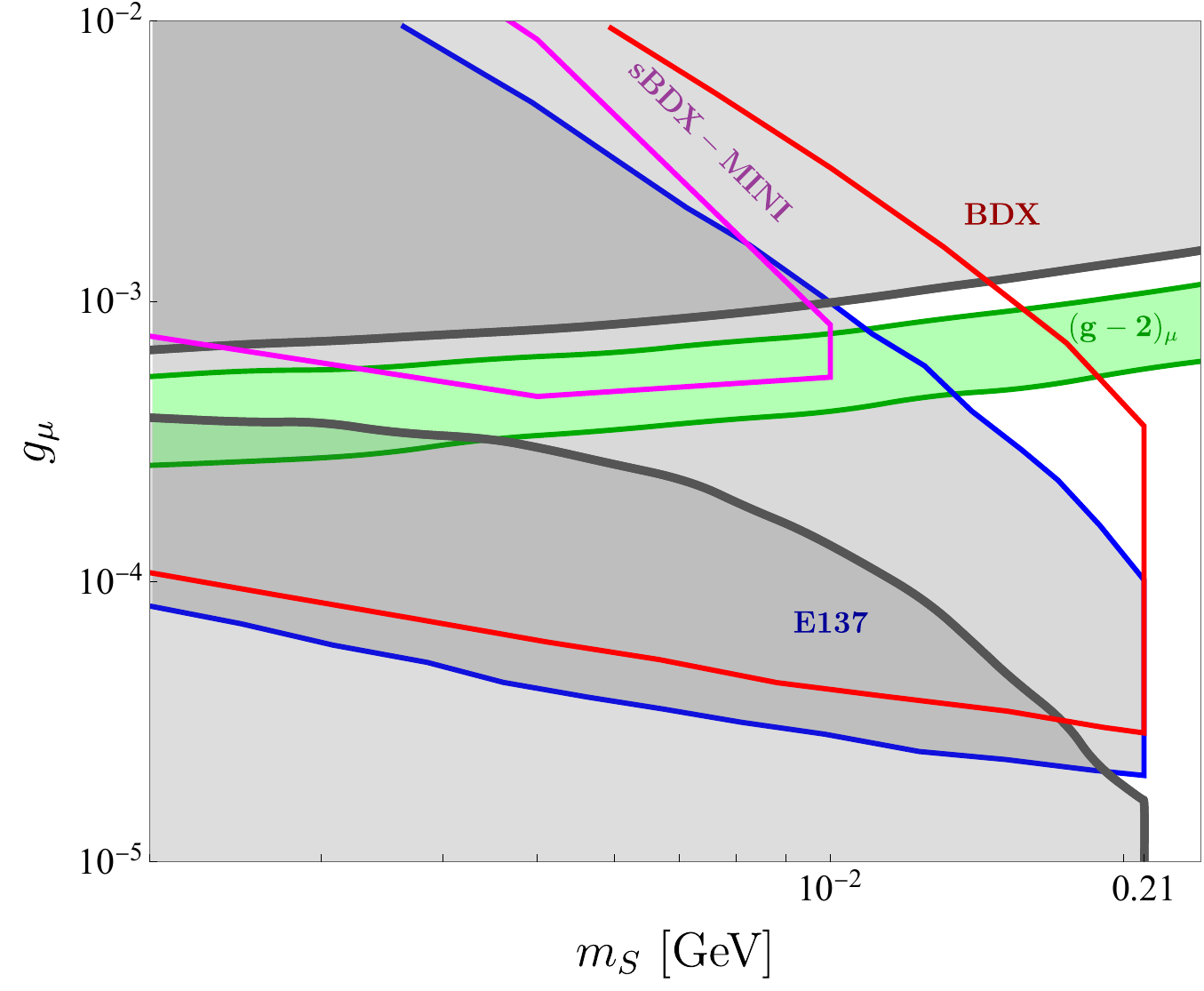}
\caption{The 90$\%$ C.L. projection of the exclusion limit of $s$BDX-MINI. E137 
exclusion 
limit (blue) and projection for BDX (red) sensitivity~\cite{Marsicano:2018vin} are also 
reported. The gray area represents already excluded regions. The green band depicts the 
parameter combinations that could explain the $(g-2)_\mu$ discrepancy. The sharp limit 
at $\sim$~0.21~GeV is related to the opening of $S\xrightarrow{}\mu\mu$ competing 
decay channel.  } 
\label{fig:scalar_reach}
\end{figure}

To evaluate the 90$\%$ C.L. exclusion limit in case of a null result, the formula $S^{UP}= 
2.3 + 1.4\sqrt{B}$~\cite{battaglieri2019dark}, where $S^{UP}$ is the upper limit on the 
number of signal events and $B$ is the total number of background events, was used. The 
expected background was conservatively estimated using BDX-MINI beam-on (at 10~GeV 
e- beam) and beam-off data ~\cite{Battaglieri:2020lds} scaled for the volume of the 
$s$BDX-MINI detector. A background yield of $\sim 0.5 \times 10^{-12}$~$\mu$/EOT is 
estimated, requiring an energy threshold of 300~MeV. The upper limit on the number of 
signal events was then translated in an exclusion limit for $g_\mu$ coupling constant. The 
exclusion limit as a function of the $S$ mass is shown in Figure~\ref{fig:scalar_reach}. 
Although $s$BDX-MINI does not test unexplored regions in the $g_\mu$~vs.~$m_S$ 
parameter space, the sensitivity that could be achieved with such a limited-size detector 
suggests that a full version of the experiment ($s$BDX) would have significant sensitivity 
to a dark scalar particle.

\section{Conclusions and Outlooks}
In this paper, we demonstrated that existing high-intensity electron beam facilities may provide low-cost, opportunistic, high-intensity secondary particle beams that will broaden their scientific programs.
We studied in detail the characteristics of muon, neutrino, and hypothetical light dark 
matter beams obtained by the interaction of the CEBAF 11~GeV primary electron beam 
with the Jefferson Lab experimental Hall-A beam dump. High statistic simulations were 
performed with the FLUKA and GEANT4 toolkit. Results show: I) a secondary muon 
beam with a Bremmstrahlung-like energy spectrum extending up to 5~GeV would provide 
up to $\sim 10^{-6}$ \textmu/EOT, corresponding to a yield of $10^8$ \textmu/s 
for an electron beam current of 50~\textmu A. II) a secondary neutrino beam with the 
typical decay-at-rest (DAR) energy spectrum would provide up to 
$\sim$$7\times10^{-5}$~$\nu$/EOT when integrated over a 1~m$^2$ detector located 
10~m above the BD. Considering a delivered charge corresponding to 10$^{22}$~EOT per 
year, the resulting integrated flux would be in the range of 10$^{18}$~$\nu$, comparable 
to dedicated flagship DAR-$\nu$ facilities such as SNS at ORNL. III) A (hypothetical) 
light dark matter \textit{leptophilic } scalar particle beam that may shed light on the 
$(g-2)_\mu$ discrepancy; this opportunity would pair with already approved 
experiments aiming to explore the dark sector, extending the BSM discovery potential of 
the Jefferson Lab.

In view of a possible upgrade to the beam energy, this study was repeated for a 22~GeV 
electron beam energy. Results showed that the CEBAF energy upgrade will be extremely 
beneficial for the secondary muon beam, extending the energy range up to 16~GeV and 
the muon flux by almost an order of magnitude. For the secondary neutrino beam, the 
DAR yield  is expected to double, and, for the dark matter beam, the dark scalar particle 
yield would increase by up to an order of magnitude.

\vspace{6pt}

\section*{Funding}
The work of C.L. and H.-S.J. is supported by the National Research Foundation of Korea (NRF) grant funded by the Korean government (MSIT) (No. NRF-2020R1F1A1077174). M.B., R.D.V.  and T.V. are part of RAISE personnel. Their work is funded by the European Union---Next Generation EU and by the Ministry of University and Research (MUR), National Recovery and Resilience Plan (NRRP), Mission 4, Component 2, Investment 1.5, project “RAISE---Robotics and AI for Socio-economic Empowerment” (ECS00000035).

\acknowledgments{We gratefully acknowledge the support of the the Jefferson Lab Radiation Control Department. We would like to thank  Jay Benesch
 for precious support to the secondary beam evaluation. The authors wish to thank also  Luca Marsicano for invaluable support to the dark matter simulation. }

\conflictsofinterest{No conflicts of interest}

\reftitle{References}

\end{document}